# Modeling and Design Optimization of a Linear Motor with Halbach Array for Semiconductor Manufacturing Technology

Sajjad Mohammadi, *Member, IEEE*, Jeffrey H. Lang, *Life Fellow,* James L. Kirtley, *Life Fellow, IEEE* and David L. Trumper, *Life Fellow, IEEE*

Massachusetts Institute of Technology, Cambridge, MA, USA

**This paper presents analytical modeling and design of a high-acceleration, low-vibration slotless double-sided linear motor with an arbitrary Halbach array for lithography machines used in semiconductor manufacturing technology. Amperian current and magnetic charge models of permanent magnets are integrated into a hybrid approach to develop comprehensive analytical modeling. Unlike conventional methods that treat magnets as sources for Poisson's equations, the solution is reduced to Laplace's equations, with magnets being represented as boundary conditions. The magnetic fields and potentials within distinct regions, along with machine quantities such as shear stress, force-angle characteristics, torque profile, attraction force, misalignment force, and back-EMF, are derived, comprehensively analyzed, and compared to FEM results for accuracy validation. In addition, two models based on Poisson's equations in terms of scalar and vector potentials are derived, compared, and analyzed. Finally, design optimization and sensitivity analysis of a linear stage for lithography applications are discussed.**

*Index Terms*—Amperian currents, FEM, Halbach array, linear motors, lithography, modeling, magnetic charges.

## I. INTRODUCTION

L INEAR PM synchronous motors have been of great interest in many industrial applications, ranging from electric trains and wave energy converters to manufacturing equipment and robotics [1]-[3]. One particular application is in lithography machines in semiconductor manufacturing technology, which demand linear motors with high acceleration and minimal vibration and acoustic noise, in which the linear motor plays a critical role in precise positioning systems to transfer a circuit pattern from a photomask onto a silicon wafer [3].

Analysis of electric motors often involves numerical techniques such as the finite element method (FEM) or other modeling frameworks. Although FEM is powerful for analyzing electric machines, from Halbach-equipped devices [4]-[5] to switched reluctance motors [6], it has drawbacks in its high cost and time-intensive nature, which impede swift designs and optimizations. On the other hand, analytical models [7]-[19] provide a deep understanding of the device's operating principles and offer profound insight into the magnetic fields and energy conversion. This is pivotal for further innovations and efficient designs. Moreover, analytical models are fast yet accurate tools, facilitating the rapid development and optimization of electromagnetic devices.

Lumped-element models, such as magnetic equivalent circuits (MEC), are highly valuable in the system-level analysis of electromechanical devices and electric machines [7]-[11]. However, they compromise accurate predictions of magnetic field distributions and precise calculations of torque ripple and back-EMF harmonics, which are critical for vibration studies [10]. This limitation becomes more pronounced when employing Halbach arrays with sophisticated field patterns [11]. An alternative approach is solving Laplace's equations [12]-[22], offering a comprehensive analytical framework with precise field solutions. The challenge lies in defining and determining boundary conditions and guessing a particular solution to address the system of equations, especially in complex geometries. Typically, PMs are modeled as sources in Poisson's equations. In [21]-[22], the solution of the diffusion equation yields eddy-currents within an actuator, which are then incorporated into a lumped-element model for designing drive and control loops. In [23], the rotor's PM is modeled as Amperian surface currents, and in [24], the magnets are modeled as magnetic charges.

The main contribution of this paper is the analytical modeling and design optimization of a slotless linear motor featuring toroidal winding and a double-sided Halbach array with arbitrary configuration for lithography machines used in semiconductor manufacturing technology, which requires low vibration and high acceleration. To provide a comprehensive design framework, various configurations, including three- and five-phase windings, as well as topologies with and without back-iron behind the PMs, are investigated. The absence of stator slots and the resulting increased effective air gap lead to smaller attraction forces and zero cogging force. The more sinusoidal field distribution of the Halbach array significantly contributes to torque ripple reduction, resulting in lower vibrations and acoustic noise. The double-sided structure not only enables higher acceleration but also cancels out normal forces and their associated harmonics, further reducing noise.

In conventional methods, an analytical model is built based on the solution of Poisson's equation with PMs represented as

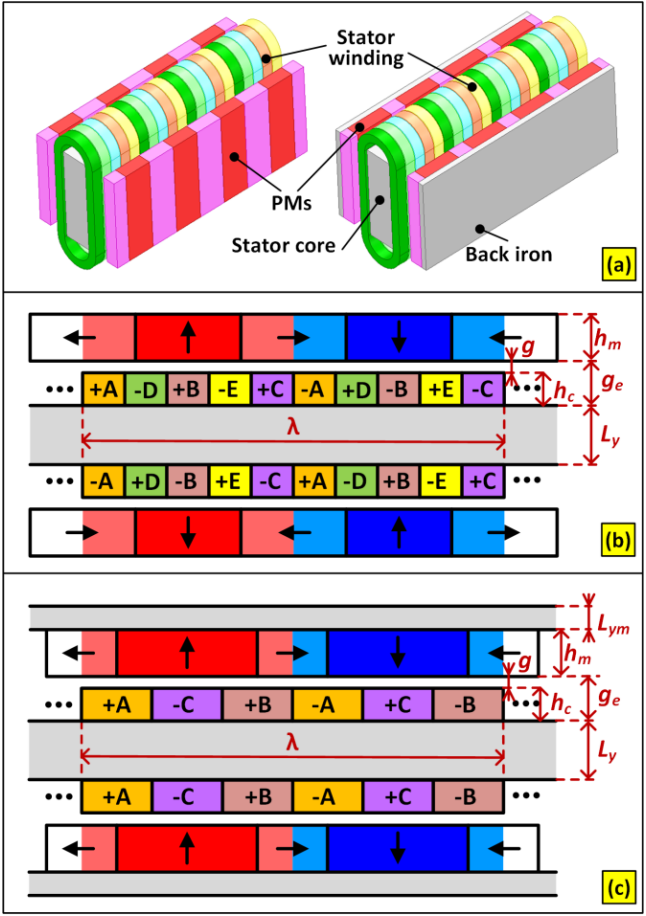

Fig. 1. Topology of the double-sided linear motor with toroidal winding and Halbach arrays: (a) 3-D view, (b) three-phase motor without back-iron behind PMs, and (c) five-phase motor without back-iron behind PMs.

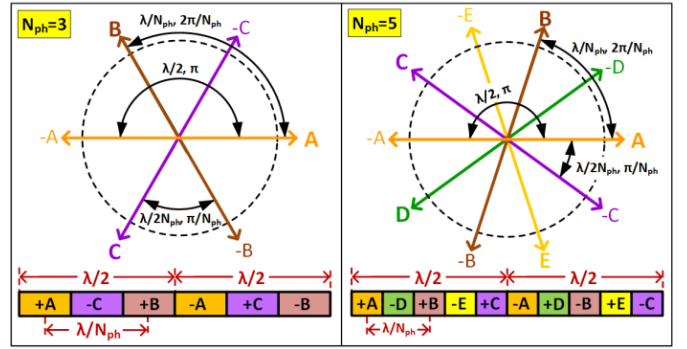

Fig. 2. Winding arrangement: (a) three-phase winding, (b) five-phase winding; Note that the signs represent the direction of winding currents

## II. MOTOR TOPOLOGY AND MODELING TECHNIQUE

Fig. 1 and Table I illustrate the topology and specifications of the linear motor under study, featuring a Halbach array and toroidal winding. The Halbach array shown here includes two PM pieces per pole ($N_m$=2). To develop a versatile model applicable across various phase windings—particularly three-phase and five-phase configurations—a three-phase winding is employed in the motor without back-iron behind PMs, commonly found, and a five-phase setup in the motor with back-iron behind PMs. Toroidal windings are employed for lower resistance and enhanced cooling efficiency compared to traditional designs. The motor is double-sided, ensuring cancellation of attraction forces on the rotor. The wave number can be expressed as below:

$$k = 2\pi / \lambda \tag{1}$$

where $\lambda$ is the pole-pair pitch, so a pole pitch is $\tau_p = \lambda/2$.

### A. Stator Winding

For a $N_{ph}$-phase winding, the current density in phase $m$ is:

$$J_m(t) = \pm J_{max} \cos(\omega t + (j-1)\frac{2\pi}{N_{ph}} + \varphi_0) \tag{2}$$

where $m$ is the sequence of the windings from left to right, $j$ is 1, 2, 3, etc., denoting for $A$, $B$, $C$, $D$ and $E$ windings, and $J_{max}$, $\omega = 2\pi f$ and $\varphi_0$ denote amplitude, frequency and initial phase of the current density, respectively. The positive or negative sign indicates the current direction, determined by the winding configuration as depicted in Fig. 2. The spatial distance and electrical angle between the positive and negative sides of each winding, as well as among the phase windings, are illustrated. The three-phase currents in the first pole pitch of the motor can be expressed as follows:

---

sources, requiring a particular solution. In the proposed approach, Poisson's equation is reduced to Laplace's equation with PMs systematically treated as boundary conditions, eliminating the need to guess any particular solution. A hybrid model for the PMs is formulated by concurrently incorporating both Amperian currents and magnetic charges, resulting in a minimal system of equations compared to the much larger system that would arise if either Amperian currents or magnetic charges were employed exclusively. Additionally, a versatile design framework is developed, allowing for any arbitrary Halbach array and winding configurations, offering substantial flexibility within the design space. Furthermore, two analytical models based on the solution of Poisson's equations in terms of scalar and vector potentials are derived, compared, and analyzed. Close-form expressions for magnetic fields and potentials within different regions, along with machine quantities such as shear stress, force-angle characteristics, force ripple, attraction force, misalignment force, and back-EMF, are derived, compared with FEM for validation, and deeply analyzed to provide profound insight for further innovations. The study delves into the impact of design parameters on the fields, shear stress, torque ripple, and back-EMF. Finally, design optimization and sensitivity analysis of a linear stage utilizing the proposed linear motor design for lithography applications is discussed.

$$J_m(t): \begin{cases} J_1(t) = J_{max}\cos(\omega t + \varphi_0); & phase\ a\ /+1 \\ J_2(t) = -J_{max}\cos(\omega t - 2\frac{2\pi}{3} + \varphi_0); & phase\ c'\ /-3 \\ J_3(t) = J_{max}\cos(\omega t - \frac{2\pi}{3} + \varphi_0); & phase\ b\ /+2 \end{cases} \quad (3)$$

The five-phase currents in the first pole pitch of the motor can be expressed as follows:

$$J_m(t): \begin{cases} J_1(t) = J_{max}\cos(\omega t + \varphi_0); & phase\ a\ /+1 \\ J_2(t) = -J_{max}\cos(\omega t - 3\frac{2\pi}{5} + \varphi_0); & phase\ d'\ /-4 \\ J_3(t) = J_{max}\cos(\omega t - 1\frac{2\pi}{5} + \varphi_0); & phase\ b\ /+2 \\ J_4(t) = -J_{max}\cos(\omega t - 4\frac{2\pi}{5} + \varphi_0); & phase\ e'\ /-5 \\ J_5(t) = J_{max}\cos(\omega t - 2\frac{2\pi}{5} + \varphi_0); & phase\ c\ /+3 \end{cases} \quad (4)$$

### B. *Halbach Array and Hybrid Modeling of PMs*

Permanent magnets are typically represented by the private solution of Poisson's equation. However, in the proposed technique, they the PMs are systematically treated as boundary conditions for Laplace's equation. This approach offers a versatile framework where any Halbach array configuration can be modeled by adjusting Fourier series coefficients in the boundary conditions. Specifically, vertical magnetizations are modeled as magnetic charges, while horizontal magnetizations are treated as Amperian currents.

### C. *Magnetic Charges for Vertical Magnetizations*

Substituting $B = \mu_0(H+M)$ into magnetic Gauss's law yields:

$$\nabla \cdot \mu_0 \vec{B} = \nabla \cdot \mu_0 (\vec{H} + \vec{M}) = 0 \implies \nabla \cdot \mu_0 \vec{H} = -\nabla \cdot \mu_0 \vec{M} \quad (5)$$

Using the Coulombian model, the PMs can be represented by magnetic charges $\rho_m$, and since the magnetization is uniform inside the PMs, there is only a surface charge density $\sigma_m$ as:

$$\rho_m = -\nabla \cdot \mu_0 \vec{M}; \quad \sigma_m = -\hat{n} \mu_0 (M^a - M^b) \quad (6)$$

where $n$ is the normal unit vector of the surface boundary. As shown in Fig. 3(a), the vectors $M$ originate from negative charges and terminates on positive charges. We obtain:

$$\sigma_m = \begin{cases} -(\pm\hat{y})\mu_0(0-M\ \hat{y}) = \pm\mu_0 M & ; top\ \&\ bottom \\ -(\pm\hat{x})\mu_0(0-M\ \hat{y}) = 0 & ; left\ \&\ right \end{cases} \quad (7)$$

where $\sigma_m$ and $-\sigma_m$ refer to the charge distribution on the bottom and top surfaces, respectively.

### D. *Amperian Currents for Horizontal Magnetizations*

By substituting $H = B/\mu_0 - M$ into Ampere's law, we have:

$$\nabla \times H = J_f \implies \nabla \times (\frac{B}{\mu_0} - M) = J_f \implies \nabla \times \frac{B}{\mu_0} = J_f + \nabla \times M \quad (8)$$

where $J_f$ is free current. A magnetization can be represented as Amperian current density $J_m$, and since it is uniform inside the PMs, there is only surface current density $K_m$ as:

$$\vec{J}_m = \nabla \times \vec{M}; \quad \vec{K}_m = \vec{M} \times \hat{n} \quad (9)$$

where $n$ is the unit vector normal to the surface of the PM. For a horizontal magnetization $M$, there are only surface currents $\pm K_m$ on the top and bottom surfaces as:

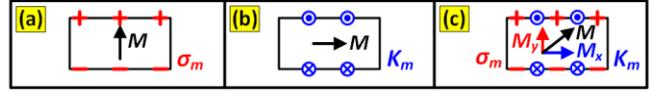

Fig. 3. Permanent magnet modeling: (a) vertical magnetization and magnetic charges, (a) horizontal magnetization and Amperian currents, and (c) angled magnetization and hybrid model.

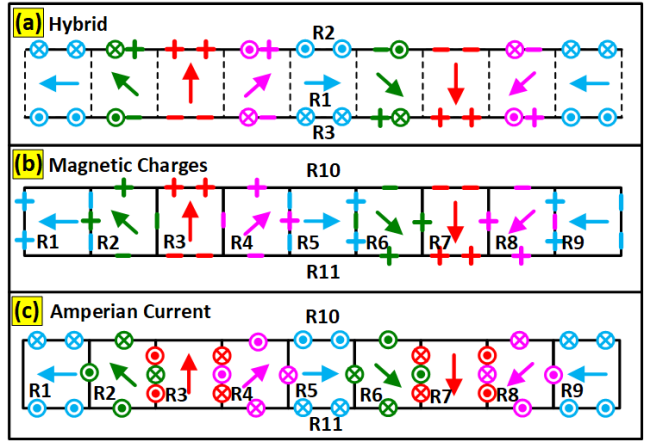

Fig. 4. Regions and boundary conditions in a Halbach array ($N_m$=4) represented by (a) hybrid model, (b) only magnetic charges, and (c) only Amperian currents.

$$\vec{K}_m = \begin{cases} \vec{M} \times (\pm\hat{y}) = M\ \hat{x} \times (\pm\hat{y}) = \pm M\ \hat{z} & ; top\ \&\ bottom \\ \vec{M} \times (\pm\hat{x}) = M\ \hat{x} \times (\pm\hat{x}) = 0 & ; left\ \&\ right \end{cases} \quad (10)$$

where $K_m$ and $-K_m$ refer to the Amperian currents at the bottom and top surfaces, respectively. As shown in Fig. 3(b), the criteria to determine the current directions is the right-hand rule. The Amperian currents are treated as free currents.

### E. *Hybrid Method for Angled Magnetizations*

As shown in Fig. 3(c), angled magnetizations are decomposed into vertical and horizontal components as:

$$\vec{M} = M_x\ \hat{x} + M_y\ \hat{y} = \overbrace{M\cos\theta^m}^{M_x}\ \hat{x} + \overbrace{M\sin\theta^m}^{M_y}\ \hat{y} \quad (11)$$

Thus, we have:

$$\begin{cases} K^m = -M_x = -M\cos\theta^m \\ \sigma^m = -\mu_0 M_y = -\mu_0 M\sin\theta^m \end{cases} \quad (12)$$

As shown in Fig. 4, the hybrid modeling approach restricts currents and charges to the horizontal boundaries at the top and bottom surfaces of the array. This simplification reduces the problem to three regions, requiring a 6-by-6 system of equations to solve. In contrast, employing either Amperian current or magnetic charges on both horizontal and vertical boundaries of each permanent magnet piece complicates the solution significantly, necessitating eleven distinct regions and a 22-by-22 system of equations to solve.

## III. HYBRID MODELING OF HALBACH ARRAYS

A characteristic of Halbach arrays is the cancellation of magnetic fields on the weaker side, which eliminates the need for a back-iron as a flux return path. Fig. 5 illustrates Halbach array configurations studied with two to five permanent magnet pieces per pole.

## A. Distribution of Magnetic Charges and Amperian Currents

The width and the electrical angular span of a PM piece is as:

$$w_m = \frac{\lambda/2}{N_m} \ , \quad \Delta\theta = \frac{\pi}{N_m} \tag{13}$$

where $N_m$ is number of PM pieces per pole. An equal rotation angle is assumed for the magnetization vectors as:

$$\Delta\theta_m = 180^o / N_m \tag{14}$$

The angle of the magnetizations in the left pole pitch is as:

$$\theta_i^m = 90^o - i\Delta\theta_m \ ; \ \ i = 0, \pm 1, \pm 2, \ldots \tag{15}$$

where $i=0$ is the angle of the vertical magnetization at the center, $i=1, 2,\ldots$ is for those at the right half of the pole pitch, and $i=-1, -2,\ldots$ is for those at the left half of the pole. The value of $i$ goes up to $(N_m\text{-}1)/2$ and $N_m/2$ for odd and even values of $N_m$, respectively. For $PM_i$, the Amperian current and the magnetic charge is as follows:

$$\begin{cases} K_i^m = -M\cos\theta_i^m \\ \sigma_i^m = -\mu_0 M\sin\theta_i^m \end{cases} \tag{16}$$

Instead of $x\epsilon[0, \lambda]$, we can go for $\theta\epsilon[0, 2\pi]$ as follows:

$$\frac{x}{\lambda} = \frac{\theta}{2\pi} \Rightarrow \theta = \frac{2\pi}{\lambda}x \Rightarrow \theta = kx \tag{17}$$

The left and right boundaries of $PM_i$ are as follows:

$$\theta_{left} = 90^o + (i-\tfrac{1}{2})\Delta\theta \ ; \ \ i = 0, \pm 1, \pm 2, \ldots \tag{18}$$

$$\theta_{right} = 90^o + (i+\tfrac{1}{2})\Delta\theta \ ; \ \ i = 0, \pm 1, \pm 2, \ldots \tag{19}$$

The Fourier series and coefficients of the Amperian current are:

$$K_m(x) = \sum_{n=1,3,\ldots}^{+\infty} k_n \cos nkx \Rightarrow K_m(\theta) = \sum_{n=1,3,\ldots}^{+\infty} k_n \cos n\theta \tag{20}$$

$$k_n = \frac{2}{\lambda}\int_0^\lambda K_m(x)\cos nkx\, dkx = \frac{2}{\pi}\int_0^\pi K_m(\theta)\cos n\theta\, d\theta \tag{21}$$

The Fourier series and coefficients of the magnetic charge are:

$$\sigma_m(x) = \sum_{n=1,3,\ldots}^{+\infty} \sigma_n \sin nkx \Rightarrow \sigma_m(\theta) = \sum_{n=1,3,\ldots}^{+\infty} \sigma_n \sin n\theta \tag{22}$$

$$\sigma_n = \frac{2}{\lambda}\int_0^\lambda \sigma_m(x)\sin nkx\, dkx = \frac{2}{\pi}\int_0^\pi \sigma_m(\theta)\sin n\theta\, d\theta \tag{23}$$

Also, the components of magnetization can be obtained as:

$$M_x(x) = -K_m(x) = \sum_{n=1,3,\ldots}^{+\infty} M_{xn}\cos nkx \Rightarrow; \ M_{xn} = -k_n \tag{24}$$

$$M_y(x) = \sum_{n=1,3,\ldots}^{+\infty} M_{yn}\sin nkx; \ M_{yn} = -\sigma_n/\mu_0 \tag{25}$$

## B. Halbach Array with Odd Number of PMs Per Pole

For odd values of $N_m$, Fourier coefficients can be determined by decomposing the integral into a summation of integrals for each piece. Amperian current coefficient can be expressed as:

$$k_n = \sum_{i=-\frac{N_m-1}{2}}^{\frac{N_m-1}{2}} \frac{2}{\pi} \int_{\frac{\pi}{2}+(i-1/2)\Delta\theta}^{\frac{\pi}{2}+(i+1/2)\Delta\theta} \overbrace{-M\cos\theta_i^m}^{K_i^m} \cos n\theta\, d\theta \tag{26}$$

$$= \sum_{i=-\frac{N_m-1}{2}}^{\frac{N_m-1}{2}} \frac{-4}{n\pi} M\cos\theta_i^m \cos[n(\frac{\pi}{2}+i\Delta\theta)]\sin[n\frac{\Delta\theta}{2}]$$

TABLE II
BOUNDARY CONDITIONS FOR THE MOTOR WITHOUT PM BACK-IRON

| | Boundary | Field | Boundary Condition |
|---|---|---|---|
| 1 | $y=0$ | $B_x$ | $H^{iron}=0 \Rightarrow H_x^I\big|_{y=0} = B_x^I\big|_{y=0} = 0$ |
| 2 | $y=g_e$ | $H_y$ | $\nabla\cdot\mu_0\vec{H}=\rho_m \Rightarrow \left(H_y^{II}-H_y^I\right)\big|_{y=g_e} = \frac{\sigma_m}{\mu_0}$ |
| 3 | | $B_x$ | $\nabla\times\frac{B}{\mu_0}=J_m \Rightarrow \left(\frac{B_x^I}{\mu_0}-\frac{B_x^{II}}{\mu_0}\right)\big|_{y=g_e} = K_m$ |
| 4 | $y=g_e+h_m$ | $H_y$ | $\nabla\cdot\mu_0\vec{H}=\rho_m \Rightarrow \left(H_y^{III}-H_y^{II}\right)\big|_{y=g_e+h_m} = -\frac{\sigma_m}{\mu_0}$ |
| 5 | | $B_x$ | $\nabla\times\frac{B}{\mu_0}=J_m \Rightarrow \left(\frac{B_x^{II}}{\mu_0}-\frac{B_x^{III}}{\mu_0}\right)\big|_{y=g_e+h_m} = -K_m$ |
| 6 | $y=+\infty$ | $B_x, H_y$ | $B_x^{III}\big|_{y=+\infty} = H_y^{III}\big|_{y=+\infty}$ |

TABLE III
BOUNDARY CONDITIONS FOR THE MOTOR WITH PM BACK-IRON

| | Boundary | Field | Boundary Condition |
|---|---|---|---|
| 1 | $y=0$ | $B_x$ | $H^{iron}=0 \Rightarrow H_x^I\big|_{y=0} = B_x^I\big|_{y=0} = 0$ |
| 2 | $y=g_e$ | $H_y$ | $\nabla\cdot\mu_0\vec{H}=\rho_m \Rightarrow \left(H_y^{II}-H_y^I\right)\big|_{y=g_e} = \frac{\sigma_m}{\mu_0}$ |
| 3 | | $B_x$ | $\nabla\times\frac{B}{\mu_0}=J_m \Rightarrow \left(\frac{B_x^I}{\mu_0}-\frac{B_x^{II}}{\mu_0}\right)\big|_{y=g_e} = K_m$ |
| 4 | $y=g_e+h_m$ | $B_x$ | $\nabla\times\frac{B}{\mu_0}=J_m \Rightarrow \left(\frac{B_x^{II}}{\mu_0}-\frac{B_x^{III}}{+\infty}\right)\big|_{y=g_e+h_m} = -K_m$ |

Magnetic charge coefficient can be expressed as:

$$\sigma_n = \sum_{i=-\frac{N_m-1}{2}}^{\frac{N_m-1}{2}} \frac{2}{\pi} \int_{\frac{\pi}{2}+(i-1/2)\Delta\theta}^{\frac{\pi}{2}+(i+1/2)\Delta\theta} \overbrace{-\mu_0 M\sin\theta_i^m}^{\sigma_i^m} \sin n\theta\, d\theta \tag{27}$$

$$= \sum_{i=-\frac{N_m-1}{2}}^{\frac{N_m-1}{2}} \frac{-4}{n\pi} M\sin\theta_i^m \sin[n(\frac{\pi}{2}+i\Delta\theta)]\sin[n\frac{\Delta\theta}{2}]$$

## C. Halbach Array with Even Number of PMs Per Pole

In the case of an even number of PM pieces per pole, the width of the first and the last PMs ($i=\pm N_m/2$) are half of that of the middle ones, so they require a separate treatment. The Fourier coefficient for Amperian currents can be calculated as follows:

$$k_n = \frac{2}{\pi}\int_0^{\frac{\pi}{2N_m}} \overbrace{-M\cos\pi}^{K_{N_m/2}^m}\cos n\theta\, d\theta + \frac{2}{\pi}\int_{\pi-\frac{\pi}{2N_m}}^{\pi} \overbrace{-M\cos 0}^{K_{N_m/2}^m}\cos n\theta\, d\theta \tag{28}$$

$$+ \sum_{i=-\frac{N_m}{2}+1}^{\frac{N_m}{2}-1} \frac{2}{\pi}\int_{\frac{\pi}{2}+(i-1/2)\Delta\theta}^{\frac{\pi}{2}+(i+1/2)\Delta\theta} \overbrace{-M\cos\theta_i^m}^{K_i^m}\cos n\theta\, d\theta$$

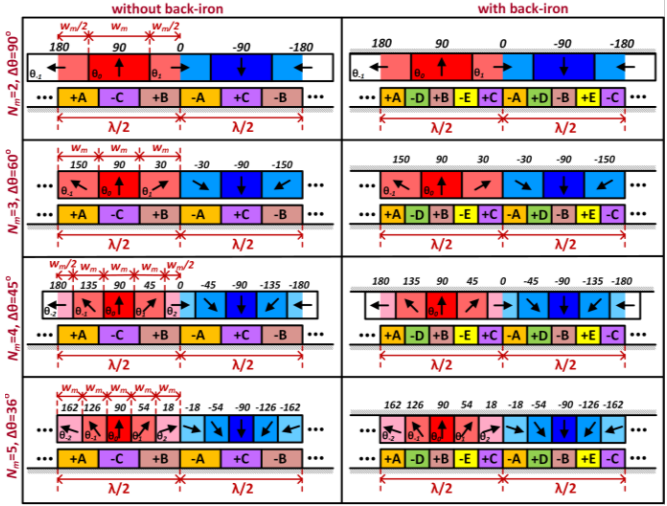

Fig. 5. Configuration of the studied linear motor without and with back-iron behind PMs for PM pieces per pole from $N_m=2$ to $N_m=5$.

$$k_n = \frac{4}{n\pi} M \sin\frac{n\pi}{2N_m} \tag{29}$$
$$+ \sum_{i=-\frac{N_m-1}{2}+1}^{\frac{N_m-1}{2}} \frac{-4}{n\pi} M \cos\theta_i^m \cos[n(\frac{\pi}{2}+i\Delta\theta)] \sin[n\frac{\Delta\theta}{2}]$$

where the first two terms pertain to the first and last pieces. The magnetic charges can be expressed as:

$$\sigma_n = \frac{2}{\pi}\int_0^{\frac{\pi}{2N_m}} \overbrace{-M\sin\pi}^{\sigma_{N_m/2}^m}\sin n\theta\, d\theta + \frac{2}{\pi}\int_{\pi-\frac{\pi}{2N_m}}^{\pi} \overbrace{-M\sin 0}^{\sigma_{N_m/2}^m}\sin n\theta\, d\theta \tag{30}$$
$$+ \sum_{i=-\frac{N_m-1}{2}+1}^{\frac{N_m-1}{2}} \frac{2}{\pi}\int_{\frac{\pi}{2}+(i-1/2)\Delta\theta}^{\frac{\pi}{2}+(i+1/2)\Delta\theta} \overbrace{-\mu_0 M\sin\theta_i^m}^{\sigma_i^m}\sin n\theta\, d\theta$$

$$\sigma_n = \sum_{i=-\frac{N_m-1}{2}+1}^{\frac{N_m-1}{2}} \frac{-4}{n\pi} M \sin\theta_i^m \sin[n(\frac{\pi}{2}+i\Delta\theta)]\sin[n\frac{\Delta\theta}{2}] \tag{31}$$

## IV. LAPLACE'S EQUATION AND BOUNDARY CONDITIONS

A notable feature

### A. Laplace's Equations in Terms of Magnetic Scalar Potential

Given that the PMs are represented as boundary conditions, Ampere's law in the *current-free regions* can be simplified to:

$$\nabla \times H = J \xrightarrow{J=0} \nabla \times H = 0 \tag{32}$$

Since the curl of the gradient of a scalar field is zero, a magnetic scalar potential can be defined as follows:

$$H = -\nabla\psi \tag{33}$$

Substitution into the magnetic Gausses' law results in the Laplace's equation as follows:

$$\nabla.B = 0 \xrightarrow{B=\mu_0 H} \nabla.\mu_0(-\nabla\psi) = 0 \rightarrow \nabla^2\psi = 0 \tag{34}$$

$$\nabla^2\psi(x,y) = \left(\frac{\partial^2\psi}{\partial x^2}+\frac{\partial^2\psi}{\partial y^2}\right) = 0 \tag{35}$$

Employing *the separation of variables* results in:

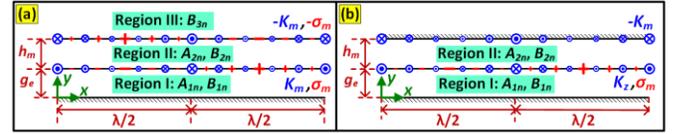

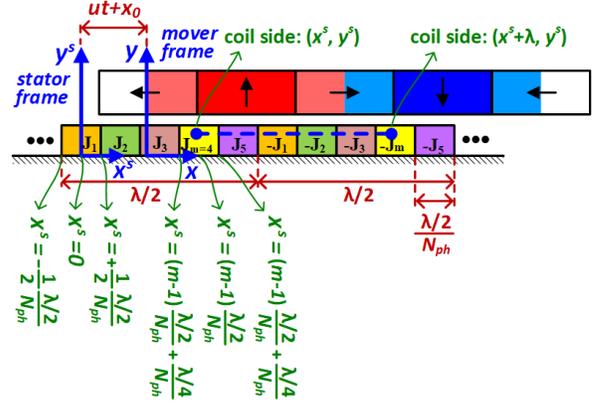

Fig. 6. Boundary conditions: (a) motor without back-iron behind PMs, and (b) motor with back-iron behind PMs.

Fig. 7. Force and back-EMF calculations, stator frame, and moving frame.

$$\psi = X(x)Y(y) \Rightarrow \frac{1}{X(x)}\frac{d^2X(x)}{dx^2} = -\frac{1}{Y(y)}\frac{d^2Y(y)}{dy^2} = k^2 \tag{36}$$

If both sides are constant and equal to a *separation constant $k^2$*, it can be simplified to a pair of ODEs as follows:

$$X'' - k^2 X = 0, \quad Y'' + k^2 Y = 0 \tag{37}$$

For $k\neq0$, since $X$ must be periodic in $x$, the solutions for $X(x)$ are the following sets

$$X(x) \sim e^{jkx}, \; e^{-jkx} \quad or \quad \sin kx, \cos kx \tag{38}$$

The exponentials are helpful for problems with infinite half-space. Also, the solutions for $Y(y)$ are as follows:

$$Y(y) \sim e^{ky}, e^{-ky} \quad or \quad \sinh ky, \cosh ky \tag{39}$$

For $k=0$, the solution for uniform fields is as follows:

$$X(x) = a_0 + b_0\, x, \quad Y(y) = 1 \tag{40}$$

However, it is not our solution as $X(x)$ should be periodic.

### B. The Regions and the General Solutions

Horizontal fields are calculated in terms of flux density $B$, determined by Amperian currents acting as boundary conditions. In contrast, vertical fields are determined in terms of field intensity $H$, where surface charge densities serve as the boundary conditions.

$$\begin{cases} B_x = -\mu_0\frac{\partial\psi}{\partial x} \Rightarrow B_{xn} = -\mu_0\frac{\partial\psi_n}{\partial x} \\ H_y = -\frac{\partial\psi}{\partial y} \Rightarrow H_{yn} = -\frac{\partial\psi_n}{\partial y} \end{cases} \tag{41}$$

At the end, we derive $H_x$ and $B_y$ as follows:

$$\begin{cases} H_x = \frac{B_x}{\mu_0} - M_x \Rightarrow H_{xn} = \frac{B_{xn}}{\mu_0} - M_{xn} \\ B_y = \mu_0(H_y + M_y) \Rightarrow B_{yn} = \mu_0(H_{yn} + M_{yn}) \end{cases} \tag{42}$$

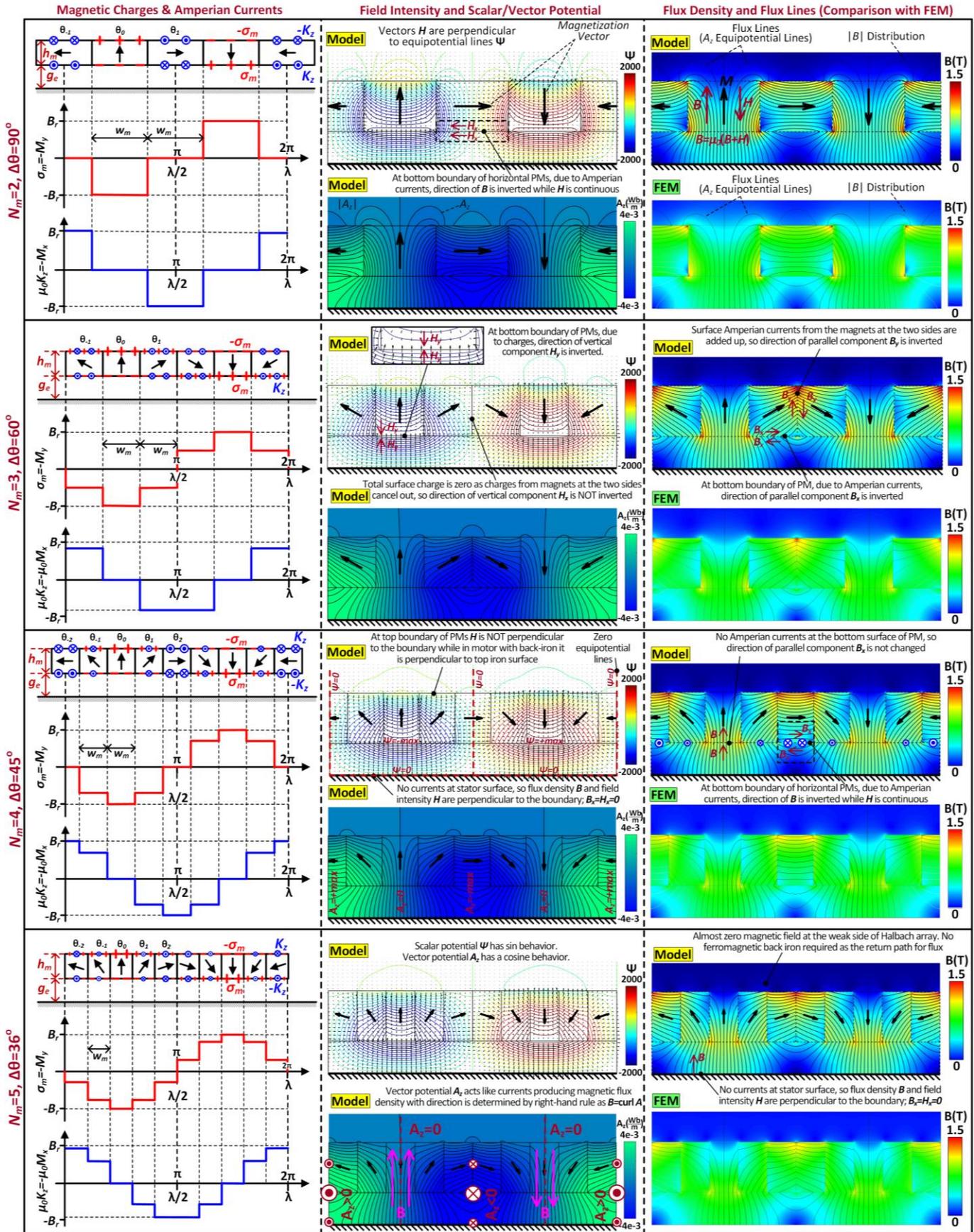

Fig. 8. Linear motor without back-iron behind PMs: (a) Amperian surfaces current and charges, (b) scalar and vector potentials, and filed intensity vectors, and (c) flux density distribution and flux lines.

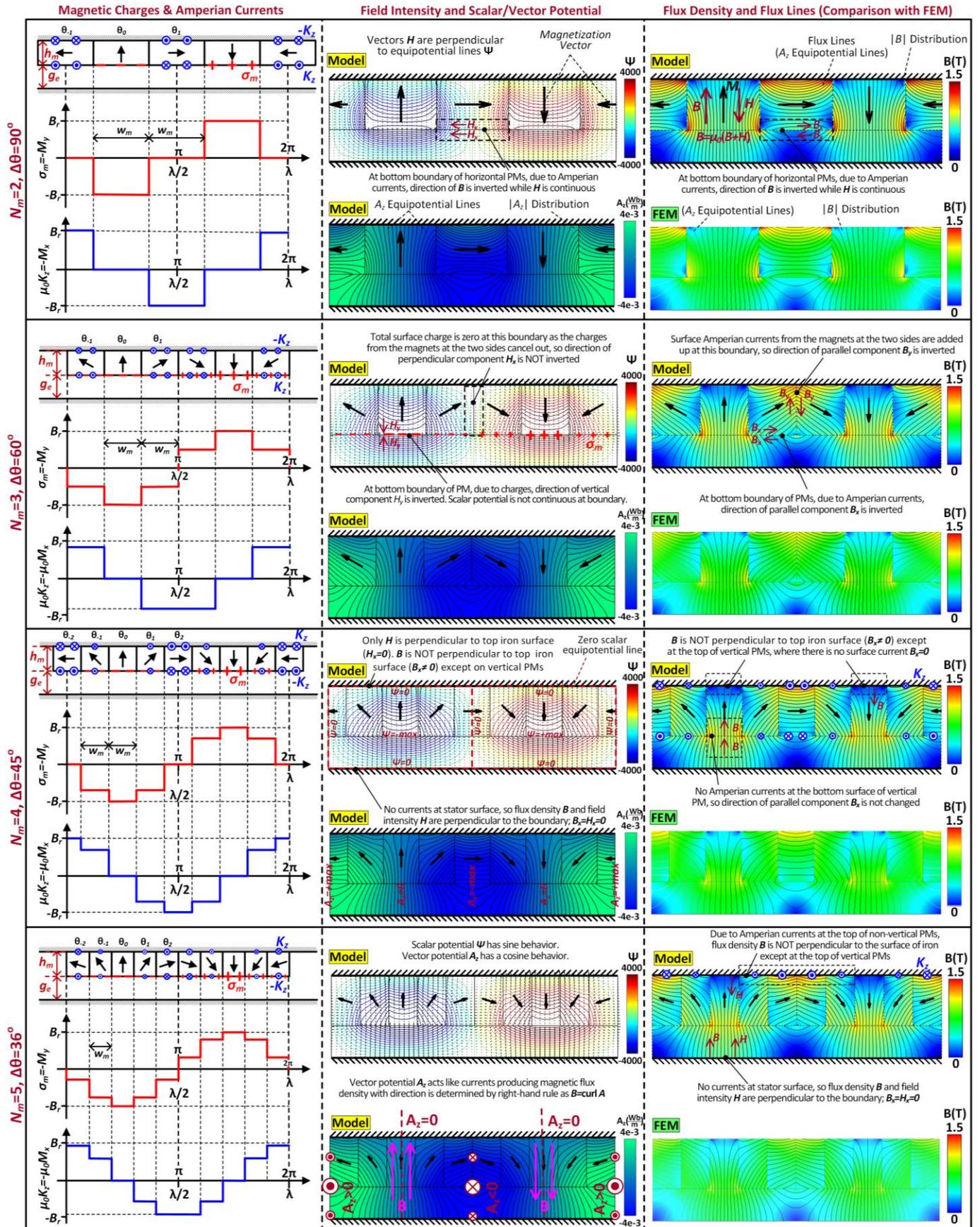

Fig. 9. Linear motor with back-iron behind PMs: (a) Amperian surfaces current and charges, (b) scalar and vector potentials, and filed intensity vectors, and (c) flux density distribution and flux lines.

simplified into three regions with boundary conditions illustrated in Fig. 6(a).

### A. The Regions and the General Solutions

The general solutions for the three regions are derived here.

#### 1. Region I: coil and airgap

The flux lines originate from the potentials at left pole ($x=\lambda/4$) and terminate on the potentials at the right pole ($x=3\lambda/4$), exemplifying an odd function of $\psi(x)$ as follows:

$$\psi^I = \sum_{n=1,3,\dots}^{+\infty} \psi_n^I = \sum_{n=1,3,\dots}^{+\infty} \overbrace{\left(A_{1n}\,e^{nky} + B_{1n}\,e^{-nky}\right)\sin nkx}^{\psi_n^I} \quad (43)$$

Then, the magnetic fields can be obtained as:

$$\begin{cases} B_x^I = \sum_{n=1,3,\dots}^{+\infty} B_{xn}^I = \sum_{n=1,3,\dots}^{+\infty} \overbrace{-\mu_0 nk\left(A_{1n}\,e^{nky} + B_{1n}\,e^{-nky}\right)\cos nkx}^{B_{xn}^I} \\ H_y^I = \sum_{n=1,3,\dots}^{+\infty} H_{yn}^I = \sum_{n=1,3,\dots}^{+\infty} \overbrace{-nk\left(A_{1n}\,e^{nky} - B_{1n}\,e^{-nky}\right)\sin nkx}^{H_{yn}^I} \end{cases} \quad (44)$$

Using the boundary condition at the surface of the stator ($A_{1n}=-B_{1n}$) that will be discussed later, it can be simplified to:

$$\begin{cases} B_x^I = \sum_{n=1,3,\dots} B_{xn}^I = \sum_{n=1,3,\dots} \overbrace{-2\mu_0 nk\,A_{1n}\sinh nky\,\cos nkx}^{B_{xn}^I} \\ H_y^I = \sum_{n=1,3,\dots} H_{yn}^I = \sum_{n=1,3,\dots} \overbrace{-2nk\,A_{1n}\cosh nky\,\sin nkx}^{H_{yn}^I} \end{cases} \quad (45)$$

#### 2. Region II: Halbach array

The general solution for the scalar potential is as follows:

$$\psi^{II} = \sum_{n=1,3,\dots}^{+\infty} \psi_n^{II} = \sum_{n=1,3,\dots}^{+\infty} \overbrace{\left(A_{2n}\,e^{nky} + B_{2n}\,e^{-nky}\right)\sin nkx}^{\psi_n^{II}} \quad (46)$$

Then the magnetic fields can be expressed as follows:

$$\begin{cases} B_x^{II} = \sum_{n=1,3,\dots}^{+\infty} B_{xn}^{II} = \sum_{n=1,3,\dots}^{+\infty} \overbrace{-\mu_0 nk\left(A_{2n}\,e^{nky} + B_{2n}\,e^{-nky}\right)\cos nkx}^{B_{xn}^{II}} \\ H_y^{II} = \sum_{n=1,3,\dots}^{+\infty} H_{yn}^{II} = \sum_{n=1,3,\dots}^{+\infty} \overbrace{-nk\left(A_{2n}\,e^{nky} - B_{2n}\,e^{-nky}\right)\sin nkx}^{H_{yn}^{II}} \end{cases} \quad (47)$$

#### 3. Region III: free space behind PMs

Requiring the potentials and fields to be finite at $y=+\infty$, the coefficient $A_{3n}$ must be zero. Consequently, the general solution for the scalar potential is expressed as follows:

$$\psi^{III} = \sum_{n=1,3,\dots}^{+\infty} \psi_n^{III} = \sum_{n=1,3,\dots}^{+\infty} \overbrace{B_{3n}\,e^{-nky}\sin nkx}^{\psi_n^{III}} \quad (48)$$

Then, the field intensities are obtained as in below:

$$\begin{cases} B_x^{III} = \sum_{n=1,3,\dots}^{+\infty} B_{xn}^{III} = \sum_{n=1,3,\dots}^{+\infty} \overbrace{-\mu_0 nk\,B_{3n}\,e^{-nky}\cos nkx}^{B_{xn}^{III}} \\ H_y^{III} = \sum_{n=1,3,\dots}^{+\infty} H_{yn}^{III} = \sum_{n=1,3,\dots}^{+\infty} \overbrace{nk\,B_{3n}\,e^{-nky}\sin nkx}^{H_{yn}^{III}} \end{cases} \quad (49)$$

### B. Boundary Conditions

To determine the values of the six unknowns $A_{1n}$, $B_{1n}$, $A_{2n}$, $B_{2n}$, $A_{3n}$ and $B_{3n}$, a set of six boundary conditions, outlined in Table II, is needed to create a system of equations.

#### 1. Tangential components at stator surface ($y=0$):

Within infinitely permeable iron, field intensity is zero, and in the absence of a surface current, the field vectors are perpendicular to the surface, implying that the tangential field $B_x$ must be null as given in Table II which leads to:

$$B_x^I\big|_{y=0} = 0 \implies A_{1n} + B_{1n} = 0 \quad (50)$$

#### 2. Normal components at boundary below PMs ($y=g_e$):

As specified in Table II, a boundary condition for $H_y$ pertain to the surface charges $\sigma_m$, yielding the following equation:

$$A_{1n}\,e^{nkg_e} - B_{1n}\,e^{-nkg_e} - A_{2n}\,e^{nkg_e} + B_{2n}\,e^{-nkg_e} = \frac{\sigma_n}{\mu_0 nk} \quad (51)$$

#### 3. Tangential components at bottom surface of PMs ($y=g_e$):

As presented in Table II, a boundary condition for $B_x$ arises from the surface currents $K_m$, yielding the subsequent equation:

$$-A_{1n}\,e^{nkg_e} - B_{1n}\,e^{-nkg_e} + A_{2n}\,e^{nkg_e} + B_{2n}\,e^{-nkg_e} = \frac{k_n}{nk} \quad (52)$$

#### 4. Normal components at top surface of PMs ($y=g_e+h_m$):

As given in Table II, a boundary condition for $H_y$ pertains to the surface charges $-\sigma_m$, yielding the following equation:

$$A_{2n}\,e^{nk(g_e+h_m)} - B_{2n}\,e^{-nk(g_e+h_m)} + B_{3n}\,e^{-nk(g_e+h_m)} = -\frac{\sigma_n}{\mu_0 nk} \quad (53)$$

#### 5. Tangential components at top surface of PMs ($y=g_e+h_m$):

As outlined in Table II, a boundary condition for $B_x$ arises from the surface currents $-K_m$, yielding the subsequent equation:

$$-A_{2n}\,e^{nk(g_e+h_m)} - B_{2n}\,e^{-nk(g_e+h_m)} + B_{3n}\,e^{-nk(g_e+h_m)} = -\frac{k_n}{nk} \quad (54)$$

#### 6. Fields at far distance ($y=+\infty$):

As outlined in Table II, to guarantee a null value for fields at infinity, the condition $A_{3n}=0$ is derived which is initially applied to the general solutions of region III to prevent introduction of an extra order to the system of equations.

### C. System of Equations

The aforementioned equations lead to the following 5-by-5 system of equations whose solution is given in the Appendix.

$$\begin{bmatrix} 1 & 1 & 0 & 0 & 0 \\ e^{nkg_e} & -e^{-nkg_e} & -e^{nkg_e} & e^{-nkg_e} & 0 \\ -e^{nkg_e} & -e^{-nkg_e} & e^{nkg_e} & e^{-nkg_e} & 0 \\ 0 & 0 & e^{nk(g_e+h_m)} & -e^{-nk(g_e+h_m)} & e^{-nk(g_e+h_m)} \\ 0 & 0 & -e^{nk(g_e+h_m)} & -e^{-nk(g_e+h_m)} & e^{-nk(g_e+h_m)} \end{bmatrix} \begin{bmatrix} A_{1n} \\ B_{1n} \\ A_{2n} \\ B_{2n} \\ B_{3n} \end{bmatrix} = \begin{bmatrix} 0 \\ \sigma_n / nk\,\mu_0 \\ k_n / nk \\ -\sigma_n / nk\,\mu_0 \\ -k_n / nk \end{bmatrix} \quad (55)$$

## VI. Linear Motor with Back-Iron Behind PMs

A back iron can still be used for magnetic isolation of the machine, structural support, and design flexibility, especially when achieving a null back field is not guaranteed due to varying widths of PM pieces. The five-phase linear motor shown in Fig. 5(b) can be simplified to two regions with boundary conditions illustrated in Fig. 6(b), where the back-iron short circuits the two poles, neutralizing positive and negative charges and leaving only surface Amperian currents along the top surface of the array.

### A. The Regions and the General Solutions

The general solutions for the two regions are derived here.

#### 1. Region I: coil and airgap

As the scalar potential is an odd function of $x$, the general solutions of potential and fields can be expressed as follows:

$$\psi^I = \sum_{n=1,3,\ldots}^{+\infty} \psi_n^I = \sum_{n=1,3,\ldots}^{+\infty} \overbrace{\left(A_{1n}\,e^{nky} + B_{1n}\,e^{-nky}\right)\sin nkx}^{\psi_n^I} \quad (56)$$

$$\begin{cases} B_x^I = \sum_{n=1,3,\ldots}^{+\infty} B_{xn}^I = \sum_{n=1,3,\ldots}^{+\infty} \overbrace{-\mu_0 nk\left(A_{1n}\,e^{nky} + B_{1n}\,e^{-nky}\right)\cos nkx}^{B_{xn}^I} \\ H_y^I = \sum_{n=1,3,\ldots}^{+\infty} H_{yn}^I = \sum_{n=1,3,\ldots}^{+\infty} \overbrace{-nk\left(A_{1n}\,e^{nky} - B_{1n}\,e^{-nky}\right)\sin nkx}^{H_{yn}^I} \end{cases} \quad (57)$$

Using the boundary condition at the surface of the stator ($A_{1n}=-B_{1n}$) that will be discussed later, it can be reduced to:

$$\begin{cases} B_x^I = \sum_{n=1,3,\ldots}^{+\infty} B_{xn}^I = \sum_{n=1,3,\ldots}^{+\infty} \overbrace{-2\mu_0 nk\,A_{1n}\sinh nky\,\cos nkx}^{B_{xn}^I} \\ H_y^I = \sum_{n=1,3,\ldots}^{+\infty} H_{yn}^I = \sum_{n=1,3,\ldots}^{+\infty} \overbrace{-2nk\,A_{1n}\cosh nky\,\sin nkx}^{H_{yn}^I} \end{cases} \quad (58)$$

#### 2. Region II: Halbach array

The general solution for the scalar potential and the magnetic fields within the Halbach array can be obtained as follows:

$$\psi^{II} = \sum_{n=1,3,\ldots}^{+\infty} \psi_n^{II} = \sum_{n=1,3,\ldots}^{+\infty} \overbrace{\left(A_{2n}\,e^{nky} + B_{2n}\,e^{-nky}\right)\sin nkx}^{\psi_n^{II}} \quad (59)$$

$$\begin{cases} B_x^{II} = \sum_{n=1,3,\ldots}^{+\infty} B_{xn}^{II} = \sum_{n=1,3,\ldots}^{+\infty} \overbrace{-\mu_0 nk\left(A_{2n}\,e^{nky} + B_{2n}\,e^{-nky}\right)\cos nkx}^{B_{xn}^{II}} \\ H_y^{II} = \sum_{n=1,3,\ldots}^{+\infty} H_{yn}^{II} = \sum_{n=1,3,\ldots}^{+\infty} \overbrace{-nk\left(A_{2n}\,e^{nky} - B_{2n}\,e^{-nky}\right)\sin nkx}^{H_{yn}^{II}} \end{cases} \quad (60)$$

### B. Boundary Conditions

To solve for the four unknowns $A_{1n}$, $B_{1n}$, $A_{2n}$, and $B_{2n}$, a set of four boundary conditions, listed in Table III, is needed.

#### 1. Tangential components at stator surface ($y=0$):

As indicated in Table III, the condition of a zero tangential field at the iron surface yields the equation below

$$B_x^I\big|_{y=0} = 0 \Rightarrow A_{1n} + B_{1n} = 0 \quad (61)$$

#### 2. Normal components at bottom surface of PMs ($y=g_e$):

As given in Table III, a boundary condition for $H_y$ associated with the surface charges $\sigma_m$ yields the following equation:

$$A_{1n}\,e^{nkg_e} - B_{1n}\,e^{-nkg_e} - A_{2n}\,e^{nkg_e} + B_{2n}\,e^{-nkg_e} = \frac{\sigma_n}{\mu_0 nk} \quad (62)$$

#### 3. Tangential components at bottom surface of PMs ($y=g_e$):

As presented in Table II, a boundary condition for $B_x$ arises from the surface currents $K_m$, yielding the subsequent equation:

$$-A_{1n}\,e^{nkg_e} - B_{1n}\,e^{-nkg_e} + A_{2n}\,e^{nkg_e} + B_{2n}\,e^{-nkg_e} = \frac{k_n}{nk} \quad (63)$$

#### 4. Tangential components at top surface of PMs ($y=g_e+h_m$):

By employing the surface current $-K_m$, and considering the null field intensity within infinitely permeable iron $H^{iron}=0$, the following equation is obtained:

$$-A_{2n}\,e^{nk(g_e+h_m)} - B_{2n}\,e^{-nk(g_e+h_m)} = -\frac{k_n}{nk} \quad (64)$$

It is interesting to note that at this boundary:

$$B_x^{II} = -\mu_0 K_m \,;\; H_x^{II} = 0 \quad (65)$$

### C. System of Equations

These four boundary conditions lead to the following 4-by-4 system of equations whose solution is given in the Appendix.

$$\begin{bmatrix} 1 & 1 & 0 & 0 \\ e^{nkg_e} & -e^{-nkg_e} & -e^{nkg_e} & e^{-nkg_e} \\ -e^{nkg_e} & -e^{-nkg_e} & e^{nkg_e} & e^{-nkg_e} \\ 0 & 0 & -e^{nk(g_e+h_m)} & -e^{-nk(g_e+h_m)} \end{bmatrix} \begin{bmatrix} A_{1n} \\ B_{1n} \\ A_{2n} \\ B_{2n} \end{bmatrix} = \begin{bmatrix} 0 \\ \sigma_n/nk\mu_0 \\ k_n/nk \\ -k_n/nk \end{bmatrix} \quad (66)$$

## VII. Forces and Back-EMF

### A. Shear Stress and Drag Force

As illustrated in Fig. 7, the field solutions and the stator current distribution are expressed in the rotor's reference frame $x$ and stator's reference frame $x^s$, respectively. The relationship connecting these two frames is as follows:

$$x^s = x + ut + x_0 \Rightarrow x = x^s - ut - x_0 \quad (67)$$

where $x_0$ is the initial position of rotor and $u$ denotes the mover velocity whose synchronous value is as follows:

$$u = 2f\tau = f\lambda \quad (68)$$

The force producing component of the rotor's field can be expressed in the stator reference frame as follows:

$$B_y^I(x^s, y^s) = \sum_{n=1,3,\ldots}^{+\infty} -2\mu_0 nk A_{1n}\cosh nky^s\,\sin nk(x^s - ut - x_0) \quad (69)$$

For a $N_{ph}$-phase motor, the current density $J_m$ within the first pole pitch with a width of $\frac{\lambda/4}{N_{ph}}$ and centered at $x^s = (m-1)\frac{\lambda/2}{N_{ph}}$ in the stator reference frame, yields a drag force as follows:

$$F_m(t) = 4L \int_0^{h_c} \int_{(m-1)\frac{\lambda/2}{N_{ph}} - \frac{\lambda/4}{N_{ph}}}^{(m-1)\frac{\lambda/2}{N_{ph}} + \frac{\lambda/4}{N_{ph}}} J_m(t) B_y^I(x^s, y^s)\,dx^s dy^s \quad (70)$$

The factor of 4 takes into consideration both sides of the phase winding in the left and right pole pitches, along with the bottom side of the stator. Substitution of flux density as detailed in the Appendix, we obtain:

$$F_m(t) = \sum_{n=1,3,\ldots}^{+\infty} \frac{-8\mu_0 L}{nk} J_m(t) A_{1n}\sinh nkh_c \times$$
$$\left\{\cos[n(m-\frac{3}{2})\frac{\pi}{N_{ph}} - nk(ut+x_0)] - \right.$$
$$\left.\cos[n(m-\frac{1}{2})\frac{\pi}{N_{ph}} - nk(ut+x_0)]\right\} \quad (71)$$

The total force produced by all phases can be obtained as:

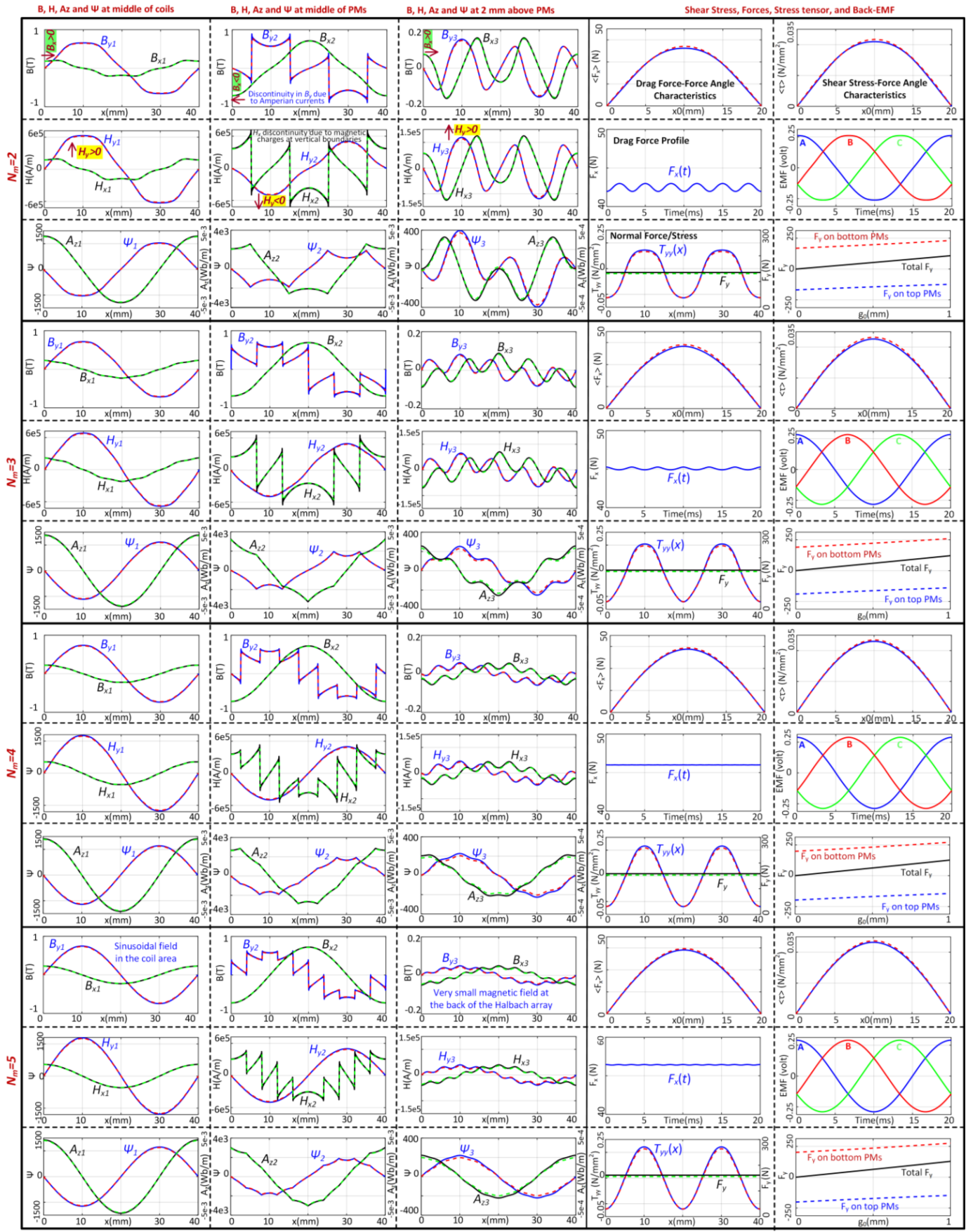

Fig. 10. Linear motor without back-iron behind PMs: flux density, field intensity, scalar potential, vector potential, force-position characteristics, force profile, misalignment force, and back-EMF (solid lines: model; dashed lines: FEM)

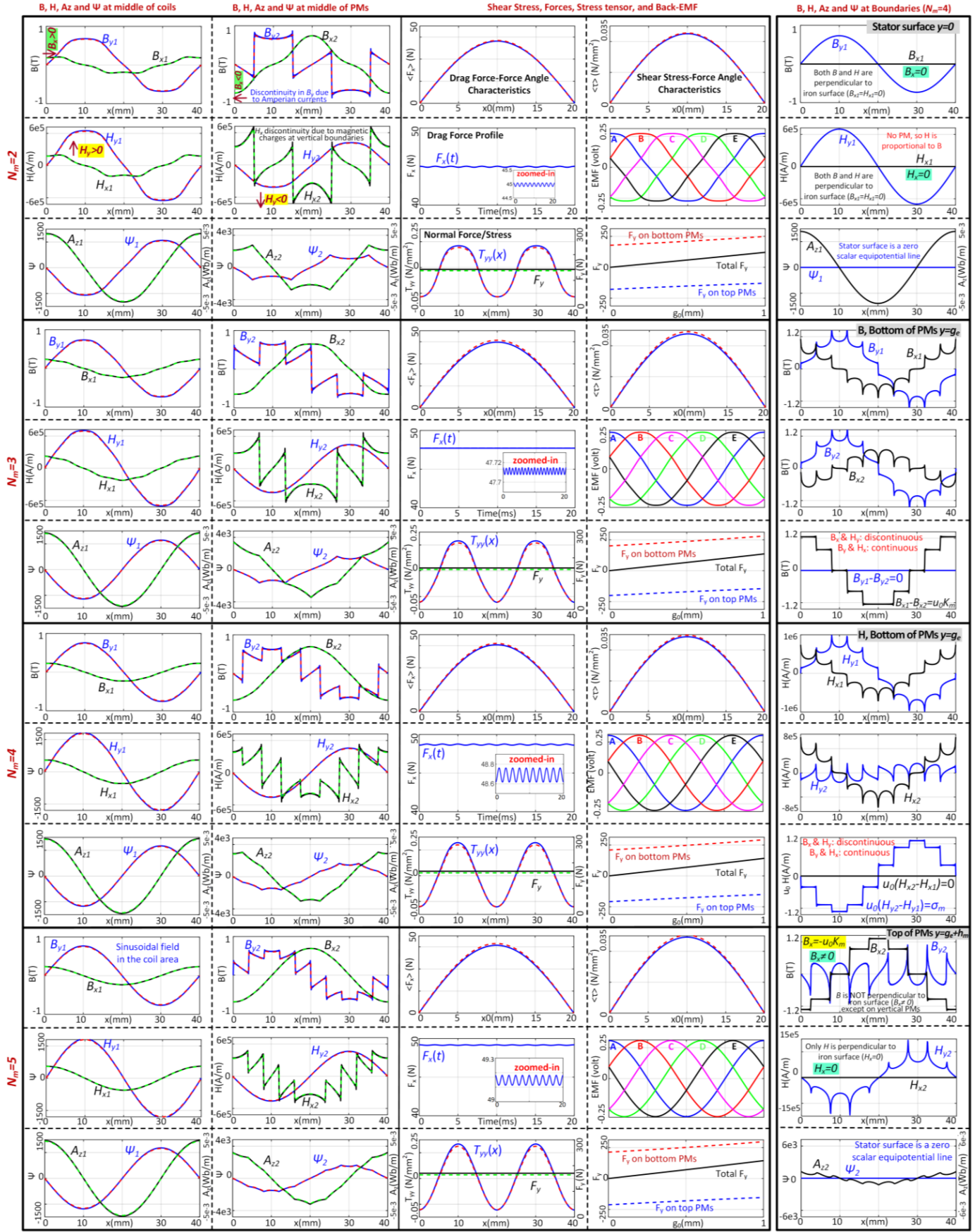

Fig. 11. Linear motor without back-iron behind PMs: flux density, field intensity, scalar potential, vector potential, force-position characteristics, force profile, misalignment force, and back-EMF (solid lines: model; dashed lines: FEM). (solid lines: model; dashed lines: FEM)

$$\tau(t) = \frac{F_t(t)}{\lambda L} \tag{72}$$

### B. Normal Stress and Attraction Force due to PMs

The normal stress or attraction force on stator per unit area due to the magnet can be determined using Maxwell stress tensors as in below:

$$T_{yy}(x, y) = \frac{(B_y^I)^2 - (B_x^I)^2}{2\mu_0} \tag{73}$$

Any value of $y$ within region I can be used, but given that $B_x$ is zero at the surface of the stator ($y=0$), it can be simplified to:

$$T_{yy}(x, y = 0) = \frac{(B_y^I\big|_{y=0})^2}{2\mu_0} = 2\mu_0 \left( \sum_{n=1,3,\dots}^{+\infty} nk A_{1n} \sin nkx \right)^2 \tag{74}$$

Thus, the total attraction force is obtained as follows:

$$F_y = \int_0^L \int_0^\lambda T_{yy}\, dx\, dz = 2\mu_0 L \sum_{n=1,3,\dots}^{+\infty} (nkA_{1n})^2 \left( \frac{\lambda}{2} - \frac{1}{4nk}\sin 2nk\lambda \right) \tag{75}$$

A mechanical misalignment with an airgap offset $g_0$ results in asymmetric airgaps and a non-zero vertical force as follows:

$$g_{top} = g - g_0 ; \quad g_{bottom} = g + g_0 \tag{76}$$

$$F_{y,total} = F_{y,top} - F_{y,bottom} \tag{77}$$

### C. Back-EMF

The back-EMFs can be obtained using the winding flux linkage along with the rotor's motion. As shown in Fig. 7, the flux linked by a coil of $m^{th}$ phase winding, with sides at $(x^s, y^s)$ and $(x^s+\lambda/2, y^s)$, can be expressed as follows:

$$\varphi_m^0(x^s, y^s) = 2\int_0^L \int_{x^s}^{x^s+\frac{\lambda}{2}} B_y^I(x^s, y^s)\, dx^s dz^s = 2L \int_{x^s}^{x^s+\frac{\lambda}{2}} B_y^I(x^s, y^s)\, dx^s dz^s \tag{78}$$

The factor of 2 is to account for the bottom side of the stator. Substituting the flux density leads to the following:

$$\varphi_m^0(x^s, y^s) = \sum_{n=1,3,\dots}^{} \frac{-8\mu_0 L A_{1n}}{nk} \cosh nky^s \cos nk(x^s - ut - x_0) \tag{79}$$

The average flux linked by a coil is determined by taking the average over the coil area as follows:

$$\varphi_m^{ave} = \frac{1}{\frac{\lambda/2}{N_{ph}}} \int_0^{h_c} \int_{(m-1)\frac{\lambda/2}{N_{ph}} - \frac{\lambda/4}{N_{ph}}}^{(m-1)\frac{\lambda/2}{N_{ph}} + \frac{\lambda/4}{N_{ph}}} \varphi_0(x^s, y^s)\, dx^s dy^s \tag{80}$$

Substituting for $\varphi_0$ yield the subsequent equation:

$$\varphi_m^{ave} = \frac{1}{\frac{\lambda/2}{N_{ph}}} \sum_{n=1,3,\dots}^{+\infty} \frac{8\mu_0 L}{n^2 k^2} A_{1n} \frac{1}{nk} \sinh nk h_c \times$$
$$\left\{ \sin[n(m-\frac{3}{2})\frac{\pi}{N_{ph}} - nk(ut+x_0)] - \right.$$
$$\left. \sin[n(m-\frac{1}{2})\frac{\pi}{N_{ph}} - nk(ut+x_0)] \right\} \tag{81}$$

With $N$ turns per coil area, the winding flux linkage and the back-EMF can be calculated as in below:

$$E_m = \frac{d\lambda_m}{dt} ; \quad \lambda_m = N \varphi_m^{ave} \tag{82}$$

Substituting for $\lambda_m$ leads to the equation below:

$$E_m(t) = \frac{N}{\frac{h_c}{\lambda/2}} \sum_{n=1,3,\dots}^{+\infty} \frac{-8\mu_0 L u}{nk} A_{1n} \sinh nk h_c$$
$$\left\{ \cos[n(m-\frac{3}{2})\frac{\pi}{N_{ph}} - nk(ut+x_0)] - \right.$$
$$\left. \cos[n(m-\frac{1}{2})\frac{\pi}{N_{ph}} - nk(ut+x_0)] \right\} \tag{83}$$

An alternative method for deriving force and verifying calculations is to use converted power as follows:

$$F_t(t) = \frac{P_t(t)}{u} = \frac{\sum_{m=1}^{N_{ph}} E_m(t) I_m(t)}{u} ; \quad I_m(t) = \frac{1}{N} \frac{h_c}{N_{ph}} \frac{\lambda/2}{N_{ph}} J_m(t) \tag{84}$$

It results in the same relationship as (72).

## VIII. Results and Analysis

In this section the field solutions and motor quantities are discussed, and design of a linear stage is discussed.

### A. Magnetic Fields

In Fig. 8 and Fig. 9, the distribution of magnetic charges, surface currents, magnetic and vector potentials, field intensity vectors, flux density vectors and flux lines derived from the model are illustrated, analyzed and validated with FEM.

In the left column, it can be seen that increasing the number of PM pieces per pole results in a more sinusoidal distribution of surface charges and currents. In the Halbach arrays with and without back-iron, the sole difference lies in the absence of surface charges on the top surface of the array in the latter case. The flux density on the weaker side of the array diminishes with higher number of PM pieces. As observed in the middle column, the magnetic scalar potential follows a sine distribution, while the magnetic vector potential has a cosine distribution as it behaves like currents generating flux.

In free-current regions, such as region II, the tangential component of the field intensity $H$ remains continuous ($\nabla \times H=0$, so $H_{1t}=H_{2t}$) while the normal component may have a discontinuity in the presence of a surface charge ($\nabla . H=\rho_m$, so $H_{1n}-H_{2n}=\sigma_m$). Alternatively, the vertical component of flux density $B$ remains continuous ($\nabla . B=0$, so $B_{1n}=B_{2n}$), while the tangential component may have a discontinuity in the presence of a surface current ($\nabla \times B=\mu_0 J_m$, so $B_{1t}-B_{2t}=\mu_0 K_m$). With no surface current, both the field intensity $H$ and flux density $B$ are perpendicular to the surface of stator ($H_t=0$). However, in the present of back-iron behind the array, only the field intensity $H$ is perpendicular to the iron surface, given the absence of magnetic charges. On the other hand, the flux density $B$ is perpendicular to the surface only at the top of the vertical PMs, where there is no Amperian current. On the bottom side of the vertical magnets, flux density $B$ is continuous, while direction of field intensity $H$ is inverted due to surface charges. Conversely, on the bottom of the horizontal magnets, the field intensity $H$ remains continuous, while the direction of flux density $B$ is inverted due to Amperian current.

Figures 10 and 11 illustrate the components of flux density ($B_x$, $B_y$), field intensity ($H_x$, $H_y$), as well as scalar and vector potentials ($\psi$, $Az$) at the middle of Regions I and II, 1 mm above the array in the Region III (for the case without back iron). This shows a strong correlation with FEM results, demonstrating a discrepancy of less than 3%. In regions I and III, there is no field discontinuity along the $x$-axis as the vertical boundaries of the PM pieces are crossed. In region II, the normal component $B_x$, remains continuous ($\nabla \cdot B = \partial Bx/\partial x=0$), while a discontinuity is observed in $B_y$ due to surface currents as shown in Fig. 4(c). Conversely, the tangential component $H_y$ remains continuous ($\nabla \times H = \partial H_y/\partial x=0$), while vertical component $H_x$ shows a discontinuity due to surface charges as shown in Fig. 4(b).

In the right column of Fig. 11, the boundary fields are examined. It is apparent that on the current-free stator surface ($y$=0), $B$ and $H$ are perpendicular to the surface, and this surface is a scalar equipotential line ($\psi$=0). However, on the back-iron surface at the top of the array ($y=g_e+h_m$), only $H_x$ is zero, while $B_x$ exhibits a step-wise distribution mirroring the surface Amperian current ($B_x=-\mu_0 K_m$ with a peak of $B_r$—an interesting observation! At the lower boundary of the Halbach array, both $B_y$ and $H_x$ display continuity ($B_{y1}-B_{y2}$=0; $H_{x1}-H_{x2}$=0), while $B_x$ and $H_y$ have a discontinuity equivalent to the surface charges and currents ($B_{x1}-B_{x2}=\mu_0 K_m$; $H_{y1}-H_{y2}=\sigma_m$). The surface of the PM back-iron is also a zero scalar equipotential line.

For comparisons with FEM, finite element analysis only provides vector potential distribution, so scalar potential is derived as in below:

$$\psi = -\int_0^x H_x(x)\,dx + \psi_0; \; \psi_0 = 0 \qquad (85)$$

### B. Forces and Back-EMF

Figs. 10 and 11 present the characteristics of drag force, shear stress versus force angle, force profile, normal stress, attraction force, misalignment vertical force, and back-EMF, showing a strong correlation with FEM simulations. As summarized in Fig. 12, it is observed that the maximum force increases with a higher number of magnets per pole $N_m$, reaching a saturation point around $N_m$=5. Additionally, the peak force of a five-phase motor exceeds that of a three-phase motor. A motor without back-iron experiences only a 3-4% reduction in peak force compared to its counterpart with back iron, while offering a lighter weight suitable for applications requiring high acceleration. The influence of back-iron on torque ripple appears negligible. The percentage of force ripple demonstrates a decreasing trend with $N_m$, with motors lacking back-iron showing higher ripple for fewer PM pieces per pole ($N_m$=2 or 3).

## IX. DESIGN OPTIMIZATION OF A LINEAR STAGE FOR PHOTOLITHOGRAPHY MACHINES

Lithography machines, also known as photolithography systems, play a critical role in semiconductor manufacturing technology within the positioning system to transfer circuit patterns from a photomask onto a silicon wafer. Linear motors are essential for the precise movement of the wafer and photomask stages. High acceleration and low vibration are among the most important design criteria for the linear motors used in these applications.

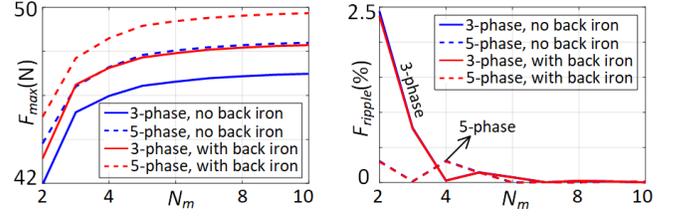

Fig. 12. Maximum force and force ripple versus number of PM pieces per pole.

In Fig. 13, an extreme ultra-violet (EUV) lithography machines (ASML NXE 3300) [3] and [25], as well as the proposed linear stage incorporating two linear motors with PMs as the moving component are illustrated. A higher drag force can be achieved by increasing the thickness of the coils or the PMs, albeit at the expense of added weight and inversely impacting acceleration. With a stage length of $L_{stage}$ and the number of units of the linear motor $N_u=L_{stage}/\lambda$, the acceleration for the moving PM and moving stator cases is obtained as follow:

$$a = 2N_u \frac{F_r}{m_{stage}+m_{Pm}}; \; m_{Pm}=4N_u\rho_{PM}\lambda L h_m \qquad (86)$$

$$a = 2N_u \frac{F_t}{m_{stage}+m_{cu}}; \; m_{cu}=4N_u\rho_{cu}\lambda L h_c \qquad (87)$$

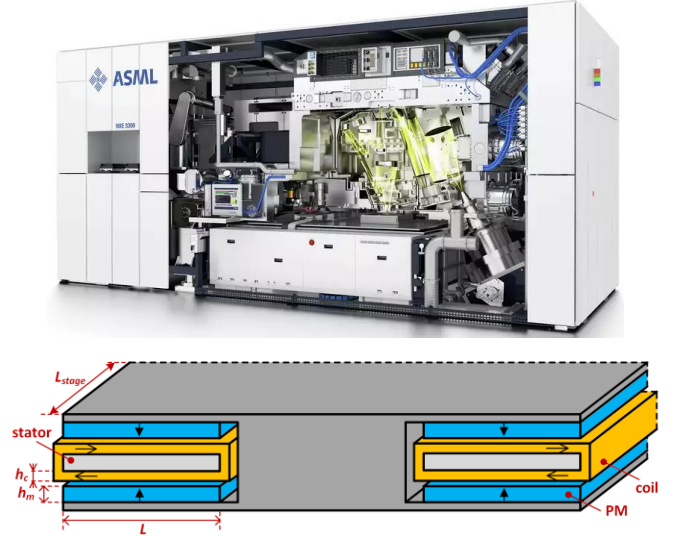

Fig. 13. ASML NXE 3300 extreme ultra-violet lithography machine (top) and the proposed moving-PM linear stage (bottom).

The the overhangs of the coils are ignored. Copper losses are approximated as in below:

$$P_{cu} = 4N_{su}\lambda L h_c J_{max}{}^2 / \sigma_{cu} \qquad (88)$$

where $N_{su}$ is the number of stator units.

According to Fig. 12, a five-phase motor without back iron, and with $N_m$=5, is employed. The sensitivity analysis of shear stress and acceleration for the linear stage, utilizing moving PMs and a moving stator, is presented in Fig. 14. Shear stress exhibits an increasing trend with $h_m$, $h_c$, and $\lambda$, eventually reaching a saturation point. On the other hand, the weight of the stage goes up with $h_m$ in the case of moving PMs and $h_c$ in the case of a moving stator, leading to a reduction in acceleration. Consequently, for the stage with moving PMs, there exists an optimal point for $h_m$ where the impact of $h_c$ on force and acceleration rise reaches a saturation point after which copper losses go up without any pay off. It represents the optimal design point for $h_m$ and $h_c$, which tends to increase for larger values of $\lambda$. Similarly, in the case of moving stator, a similar trend is observed for $h_m$, i.e. an optimum value for $h_m$ while $h_c$ reaches a saturation after which it is a waste of resources.

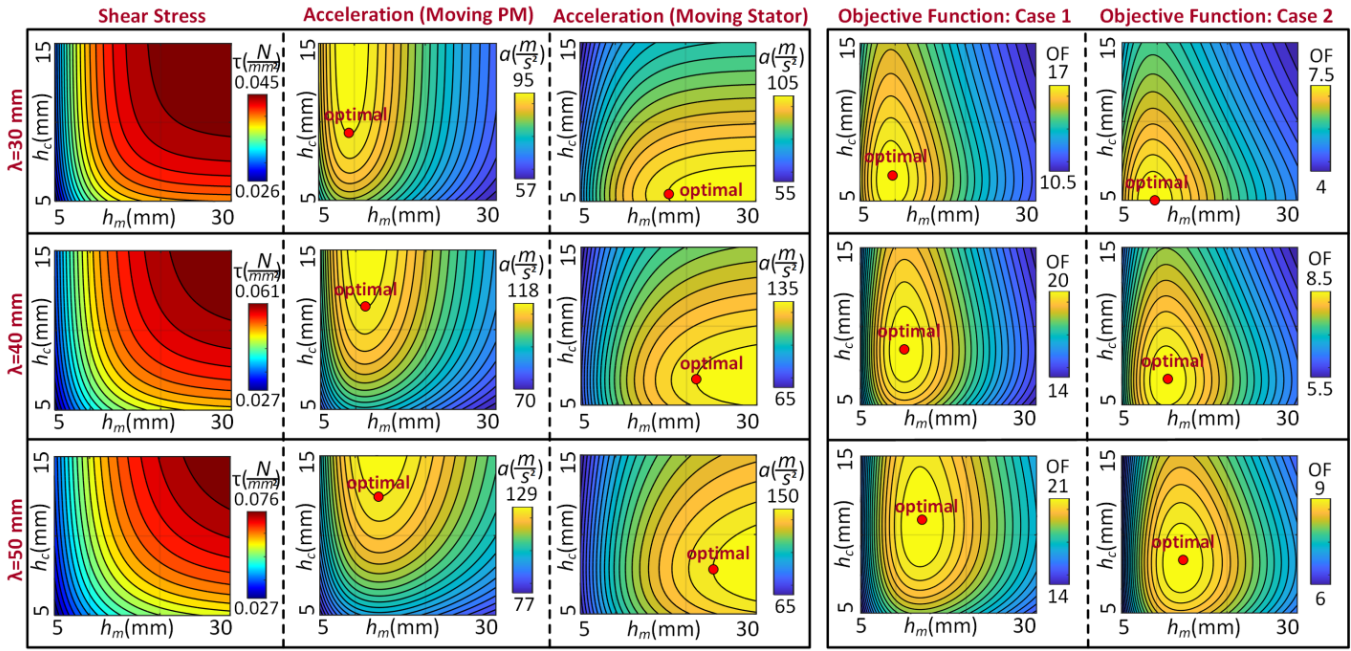

Fig. 14. Sensitivity of shear stress, acceleration of moving PM and moving stator stages, and the objective function versus $\lambda$, $h_m$ and $h_c$..

Multi objective optimization can be carried out in the design of the stage. For example, in the case of moving PM, copper loss linearly goes up with the coil height $h_c$ although it has an increasing impact on acceleration. Therefore, the following objective function can be employed for design optimizations using any optimization technique such as genetic algorithm [26]:

$$OF = \frac{a^\alpha}{P_{cu}^\beta} \qquad (89)$$

where $\alpha$ and $\beta$ are set based on the order of importance of each objective. In Fig. 14, sensitivity of $OF$ for two cases with [$\alpha$=1, $\beta$=0.2] and [$\alpha$=1, $\beta$=0.3] is shown. It can be seen that, for any value of $\lambda$, there are optimum values for PM height $h_m$ and coil height $h_c$ because, although a higher force is produced at higher values of $h_m$ and $h_c$, they have an adverse impact on acceleration and copper loss. The larger the unit length $\lambda$, the higher the optimal values of $h_m$ and $h_c$. It is also seen that, in case 2 with higher emphasize on the reduction of copper loss, a lower value is obtained for the optimal coil height $h_c$. it is worth noting that other objective such as minimization of torque ripple, back-EMF THD, and the cost of the drive can be included into the objective function as follows:

$$OF = \frac{a}{P_{cu} \times THD_E \times T_{ripple} \times Cost_{drive}} \qquad (90)$$

## X. CONCLUSION

This paper presents the design and modeling of a linear motor for lithography applications in semiconductor manufacturing technology. By integrating both Amperian current and magnetic charge models of permanent magnets, an analytical model is developed that can incorporate any number of phases and any arbitrary Halbach configuration, with or without back iron, making it a versatile and flexible toolkit for design optimizations. The proposed model offers several advantages, reducing the complexity of the solution and the computational burden without compromising accuracy. This makes it well-suited for parametric analysis and optimization algorithms, providing results within seconds for single runs and within a few minutes for optimizations. In contrast, FEM requires over an hour for a single run with an ordinary mesh, and optimizations involving numerous runs may extend to several days.

To develop the proposed model, Poisson's equation is first reduced to Laplace's equation without requiring a particular solution. Secondly, permanent magnets are employed as boundary conditions rather than sources. Thirdly, the hybrid technique necessitates the solution for only two or three regions, a notable reduction compared to the large number of regions required if either Amperian currents or magnetic charges were utilized exclusively. Motor quantities, from fields and potentials to forces and back-EMF, are analytically derived, thoroughly scrutinized to provide deep insight for future designers, and successfully validated with FEM results, demonstrating a discrepancy of less than 3%. The impact of motor parameters on maximum torque and torque ripple is studied. Finally, sensitivity analysis and design optimization for a linear stage for photolithography applications are discussed. An objective function for multi-objective design optimization is introduced.

Additionally, two analytical models using the solution of Poisson's equation in terms of scalar and vector potentials are derived, compared, and analyzed to provide further insight for future innovations. The designers may pick any of these analytical frameworks based on the problem and the geometry of the electromechanical device. The complexity of solutions based on Laplace's and Poisson's equations are almost the same. However, the Poisson's equation requires a particular solution to be guessed which can be challenging in complicated geometries and magnetization distributions. The proposed model does not require guessing a particular solution and systematically incorporates the impact of the magnetizations into the boundary conditions.

## Acknowledgement

The authors would like to thank Steve Roux and Nate Finney at ASML, as well as the members of the Research Laboratory for Electromagnetic and Electronic Systems (LEES) and the Precision Motion Control Laboratory (PMCL) at MIT for their productive discussions throughout this research.

## Appendix I: Field Coefficients

The coefficients $A_{1n}$, $B_{1n}$, $A_{2n}$, $B_{2n}$ and $B_{3n}$ for the motor without back-iron behind PMs are obtained as follows:

$$A_{1n} = \frac{1}{2\mu_0 nk}\left\{e^{-nkg_e} - e^{-nk(g_e+h_m)}\right\}\left\{\sigma_n - \mu_0 k_n\right\} \quad (A1)$$

$$B_{1n} = -A_{1n} = \frac{-1}{2\mu_0 nk}\left\{e^{-nkg_e} - e^{-nk(g_e+h_m)}\right\}\left\{\sigma_n - \mu_0 k_n\right\} \quad (A2)$$

$$A_{2n} = \frac{-1}{2\mu_0 nk}\left\{e^{-nk(g_e+h_m)}\right\}\left\{\sigma_n - \mu_0 k_n\right\} \quad (A3)$$

$$B_{2n} = \frac{k_n}{2nk}\left\{2\cosh nkg_e - e^{-nk(g_e+h_m)}\right\} \\ + \frac{\sigma_n}{2\mu_0 nk}\left\{2\sinh nkg_e + e^{-nk(g_e+h_m)}\right\} \quad (A4)$$

$$B_{3n} = \frac{\sigma_n}{\mu_0 nk}\left\{\sinh nkg_e - \sinh nk(g_e+h_m)\right\} \\ + \frac{k_n}{nk}\left\{\cosh nkg_e - \cosh nk(g_e+h_m)\right\} \quad (A5)$$

The coefficients $A_{1n}$, $B_{1n}$, $A_{2n}$, $B_{2n}$ and $B_{3n}$ for the motor without back-iron behind PMs are derived as follows:

$$A_{1n} = \frac{1}{2\mu_0 nk}\left\{\sigma_n \frac{\sinh nkh_m}{\sinh nk(g_e+h_m)} + \mu_0 k_n \frac{1 - \cosh nkh_m}{\sinh nk(g_e+h_m)}\right\} \quad (A6)$$

$$B_{1n} = -A_{1n} \quad (A7)$$

$$A_{2n} = \frac{-1}{2\mu_0 nk}\left\{\sigma_n \frac{e^{-nk(g_e+h_m)}\sinh nkg_e}{\sinh nk(g_e+h_m)} + \mu_0 k_n \frac{e^{-nk(g_e+h_m)}\cosh nkg_e - 1}{\sinh nk(g_e+h_m)}\right\} \quad (A8)$$

$$B_{2n} = \frac{1}{2\mu_0 nk}\left\{\sigma_n \frac{e^{-nk(g_e+h_m)}\sinh nkg_e}{\sinh nk(g_e+h_m)} + \mu_0 k_n \frac{e^{nk(g_e+h_m)}\cosh nkg_e - 1}{\sinh nk(g_e+h_m)}\right\} \quad (A9)$$

## Appendix II: Force Calculations

Substitution of (69) into (70) leads to the equation below:

$$F_m(t) = 4L\int_0^{h_e}\int_{(m-1)\frac{\lambda/2}{N_{pm}}-\frac{\lambda/4}{N_{pm}}}^{(m-1)\frac{\lambda/2}{N_{pm}}+\frac{\lambda/4}{N_{pm}}} J_m(t) \times \\ \left(\sum_{n=1,3,...}^{+\infty} -2\mu_0 nk \, A_{1n}\cosh nky^s \, \sin nk(x^s - ut - x_0)\right)dx^s dy^s \quad (A10)$$

$$F_m(t) = \sum_{n=1,3,...}^{+\infty}\left\{-8\mu_0 nkLJ_m(t)A_{1n}\int_0^{h_e}\cosh nky^s \, dy^s \\ \int_{(m-1)\frac{\lambda/2}{N_{pm}}-\frac{\lambda/4}{N_{pm}}}^{(m-1)\frac{\lambda/2}{N_{pm}}+\frac{\lambda/4}{N_{pm}}} \sin nk(x^s - ut - x_0)\,dx^s\right\} \quad (A11)$$

## Appendix III: Field Solution Using Poisson's Equation in Terms of Magnetic Scalar Potential (Model 2)

As summarized and compared in Tables IV and V, in addition to the model 1 which is presented earlier in the paper, two models are obtained based on the solution of Poisson's equation in terms of scalar and vector potentials. It is seen that Model 1 is developed based on a homogeneous solution while models 2 and 3 require a particular solution. Models 2 and 3 are respectively developed upon field intensity ($H_x$, $H_y$) and flux density ($B_x$, $B_y$), while model 1 takes a hybrid approach ($B_x$, $H_y$).

As in Model 2, field solutions can be obtained using Poisson's equation in terms of the magnetic scalar potential. In *current-free regions*, a magnetic scalar potential can be defined as follows:

$$\nabla \times H = J \xrightarrow{J=0} \nabla \times H = 0 \Rightarrow H = -\nabla\psi \quad (A12)$$

Substitution into the magnetic Gausses' law results in the Laplace's equation as follows:

$$\nabla.B = 0 \xrightarrow{B=\mu_0(H+M)} \nabla.\mu_0(-\nabla\psi + M) = 0 \Rightarrow \nabla^2\psi = \nabla.M \quad (A13)$$

It requires a homogeneous solution $\psi_h$ and a particular solution $\psi_p$ as follows:

$$\psi = \psi_h + \psi_p \quad (A14)$$

The homogeneous solution solves the Laplacian equation by employing *the separation of variables* as follows:

$$\nabla^2\psi_h = 0 \; ; \; \psi_h(x,y) = X'(x)Y'(y) \quad (A15)$$

where the general solutions are obtained as follows:

$$\begin{cases} X'(x) \sim e^{jkx}, \, e^{-jkx} \quad or \quad \sin kx, \cos kx \\ Y'(y) \sim e^{ky}, \, e^{-ky} \quad or \quad \sinh ky, \cosh ky \end{cases} \quad (A16)$$

The homogeneous solution for all three regions is as follows:

$$\psi_{n,h} = \left(A_n e^{nky} + B_n e^{-nky}\right)\sin nkx \quad (A17)$$

The particular solution is zero except within the Halbach array, where we need to guess a solution that satisfies the Poisson's equation below:

$$\nabla^2\psi_p = \nabla.M = \frac{\partial M_x}{\partial x} + \frac{\partial M_y}{\partial y} \quad (A18)$$

It leads to:

$$\frac{\partial^2\psi_p}{\partial x^2} + \frac{\partial^2\psi_p}{\partial y^2} = \sum_{n=1,3,...}^{+\infty} -nkM_{xn}\sin nkx \quad (A19)$$

TABLE IV
Motor without Back-Iron Behind PMs: Comparison of Solution and Boundary Conditions for The Three Models

| | | Model 1 | Model 2 | Model 3 |
|---|---|---|---|---|
| **Solution** | | Laplace's Equation: Scalar Potential $\nabla^2 \psi = 0$ | Poisson's Equation: Scalar Potential $\nabla^2 \psi = \nabla \cdot M$ | Poisson's Equation: Vector potential $\nabla^2 A_z = -\mu_0 \nabla \times M$ |
| **Homogenous Solution** | | $\psi_{n,h} = \left(A_n e^{nky} + B_n e^{-nky}\right)\sin nkx$ | $\psi_{n,h} = \left(A_n e^{nky} + B_n e^{-nky}\right)\sin nkx$ | $A_{zn,h} = \left(A_n e^{nky} + B_n e^{-nky}\right)\cos nkx$ |
| **Particular Solution** | | Not needed | $\psi_{n,p} = \dfrac{M_{xn}}{nk}\sin nkx$ | $A_{z,p} = \dfrac{\mu_0 M_{yn}}{nk}\cos nkx$ |
| **Solution in Terms of** | | Hybrid: $B_x$, $H_y$ $B_x = -\mu_0 \partial \psi / \partial x$ $H_y = -\partial \psi / \partial y$ | Field Intensity: $H_x$, $H_y$ $H = -\nabla \psi \rightarrow \begin{cases} H_x = -\partial \psi / \partial x \\ H_y = -\partial \psi / \partial y \end{cases}$ | Flux Density: $B_x$, $B_y$ $B = \nabla \times A \rightarrow \begin{cases} B_x = \partial A_z / \partial y \\ B_y = -\partial A_z / \partial x \end{cases}$ |
| **Potentials and Fields** | **Region I** | $\psi_n^I = \left(A_{1n} e^{nky} + B_{1n} e^{-nky}\right)\sin nkx$ $B_{xn}^I = -\mu_0 nk\left(A_{1n} e^{nky} + B_{1n} e^{-nky}\right)\cos nkx$ $H_{yn}^I = -nk\left(A_{1n} e^{nky} - B_{1n} e^{-nky}\right)\sin nkx$ | $\psi_n^I = \left(A'_{1n} e^{nky} + B'_{1n} e^{-nky}\right)\sin nkx$ $H_{xn}^I = -nk\left(A'_{1n} e^{nky} + B'_{1n} e^{-nky}\right)\cos nkx$ $H_{yn}^I = -nk\left(A'_{1n} e^{nky} - B'_{1n} e^{-nky}\right)\sin nkx$ | $A_{zn}^I = \left(A'_{1n} e^{nky} + B'_{1n} e^{-nky}\right)\cos nkx$ $B_{xn}^I = nk\left(A'_{1n} e^{nky} - B'_{1n} e^{-nky}\right)\cos nkx$ $B_{yn}^I = nk\left(A'_{1n} e^{nky} + B'_{1n} e^{-nky}\right)\sin nkx$ |
| | **Region II** | $\psi_n^{II} = \left(A_{2n} e^{nky} + B_{2n} e^{-nky}\right)\sin nkx$ $B_{xn}^{II} = -\mu_0 nk\left(A_{2n} e^{nky} + B_{2n} e^{-nky}\right)\cos nkx$ $H_{yn}^{II} = -nk\left(A_{2n} e^{nky} - B_{2n} e^{-nky}\right)\sin nkx$ | $\psi_n^{II} = \left(A'_{2n} e^{nky} + B'_{2n} e^{-nky} + \dfrac{M_{xn}}{nk}\right)\sin nkx$ $H_{xn}^{II} = -nk\left(A'_{2n} e^{nky} + B'_{2n} e^{-nky} + \dfrac{M_{xn}}{nk}\right)\cos nkx$ $H_{yn}^{II} = -nk\left(A'_{2n} e^{nky} - B'_{2n} e^{-nky}\right)\sin nkx$ | $A_{zn}^{II} = \left(A'_{2n} e^{nky} + B'_{2n} e^{-nky} + \dfrac{\mu_0 M_{yn}}{nk}\right)\cos nkx$ $B_{xn}^{II} = nk\left(A'_{2n} e^{nky} - B'_{2n} e^{-nky}\right)\cos nkx$ $B_{yn}^{II} = nk\left(A'_{2n} e^{nky} + B'_{2n} e^{-nky} + \dfrac{\mu_0 M_{yn}}{nk}\right)\sin nkx$ |
| | **Region III** | $\psi_n^{III} = B_{3n} e^{-nky}\sin nkx$ $B_{xn}^{III} = -\mu_0 nk\, B_{3n} e^{-nky}\cos nkx$ $H_{yn}^{III} = nk\, B_{3n} e^{-nky}\sin nkx$ | $\psi_n^{III} = B'_{3n} e^{-nky}\sin nkx$ $H_{xn}^{III} = -nk\, B'_{3n} e^{-nky}\cos nkx$ $H_{yn}^{III} = nk\, B'_{3n} e^{-nky}\sin nkx$ | $A_{zn}^{III} = B'_{3n} e^{-nky}\cos nkx$ $B_{xn}^{III} = -nk B'_{3n} e^{-nky}\cos nkx$ $B_{yn}^{III} = nk B'_{3n} e^{-nky}\sin nkx$ |
| **Boundary Condition: Motor without Back-Iron** | **Stator** | Zero: $y = 0 \rightarrow B_x^I = 0$ | Zero: $y = 0 \rightarrow H_x^I = 0$ | Zero: $y = 0 \rightarrow B_x^I = 0$ |
| | **y-Field** | Discontinuity $= \sigma_m$ $\nabla \cdot \mu_0 \vec{H} = \rho_m$ $g = g_e \rightarrow H_y^{II} - H_y^I = \sigma_m / \mu_0$ $g = g_e + h_m \rightarrow H_y^{III} - H_y^{II} = -\sigma_m / \mu_0$ | Discontinuity $= M_y$ $\nabla \cdot B = 0 \rightarrow \nabla \cdot H = -\nabla \cdot M$ $g = g_e \rightarrow \mu_0 H_y^I = \overbrace{\mu_0\left(H_y^{II} + M_y\right)}^{B_y^{II}}$ $\rightarrow H_y^I - H_y^{II} = M_y$ | No discontinuity $\nabla \cdot B = 0$ $g = g_e \rightarrow B_y^I = B_y^{II}$ $g = g_e + h_m \rightarrow B_y^{II} = B_y^{III}$ |
| | **x-Field** | Discontinuity $= K_m$ $\nabla \times (B / \mu_0) = J_m$ $g = g_e \rightarrow B_x^I - B_x^{II} = \mu_0 K_m$ $g = g_e + h_m \rightarrow B_x^{II} - B_x^{III} = -\mu_0 K_m$ | No discontinuity $g = g_e + h_m \rightarrow \nabla \times (H + M) = \overbrace{\mu_0 H_y^{III}}^{B_y^{III}}$ $\rightarrow H_y^{II} - H_y^{III} = -M_y$ $g = g_e \rightarrow H_x^I = H_x^{II}$ $g = g_e + h_m \rightarrow H_x^{II} = H_x^{III}$ | Discontinuity $= M_x$ $\nabla \times H = 0 \rightarrow \nabla \times B = \mu_0 \nabla \times M$ $g = g_e \rightarrow \overbrace{\dfrac{B_x^I}{\mu_0}}^{H_x^I} = \overbrace{\dfrac{B_x^{II}}{\mu_0}}^{H_x^{II}} - M_x$ $\rightarrow B_x^I - B_x^{II} = -\mu_0 M_x$ $g = g_e + h_m \rightarrow \overbrace{\dfrac{B_x^{II}}{\mu_0} - M_x}^{H_x^{II}} = \overbrace{\dfrac{B_x^{III}}{\mu_0}}^{H_x^{III}}$ $\rightarrow B_x^{II} - B_x^{III} = \mu_0 M_x$ |
| | **$+\infty$** | Zero | Zero | Zero |

Particular solution is not unique, and any solution satisfying the above equation works. Assuming a particular solution that is only a function of $x$, after mathematical manipulations, we can guess the following solution that works:

$$\psi_{n,p} = \sum_{n=1,3,\dots}^{+\infty} \frac{M_{xn}}{nk}\sin nkx \qquad (A20)$$

Finally, the general solution for Halbach array region is obtained as follows:

TABLE V
MOTOR WITH BACK-IRON BEHIND PMs: COMPARISON OF SOLUTION AND BOUNDARY CONDITIONS FOR THE THREE MODELS

| | | Model 1 | Model 2 | Model 3 |
|---|---|---|---|---|
| *Solution* | | Laplace's Equation: Scalar Potential $\nabla^2 \psi = 0$ | Poisson's Equation: Scalar Potential $\nabla^2 \psi = \nabla \cdot M$ | Poisson's Equation: Vector potential $\nabla^2 A_z = -\mu_0 \nabla \times M$ |
| *Homogenous Solution* | | $\psi_{n,h} = \left(A_n e^{nky} + B_n e^{-nky}\right)\sin nkx$ | $\psi_{n,h} = \left(A_n e^{nky} + B_n e^{-nky}\right)\sin nkx$ | $A_{zn,h} = \left(A_n e^{nky} + B_n e^{-nky}\right)\cos nkx$ |
| *Particular Solution* | | Not needed | $\psi_{n,p} = \dfrac{M_{xn}}{nk}\sin nkx$ | $A_{z,p} = \dfrac{\mu_0 M_{yn}}{nk}\cos nkx$ |
| *Solution in Terms of* | | Hybrid: $B_x$, $H_y$ $$B_x = -\mu_0 \partial\psi / \partial x$$ $$H_y = -\partial\psi / \partial y$$ | Field Intensity: $H_x$, $H_y$ $$H = -\nabla\psi \to \begin{cases} H_x = -\partial\psi/\partial x \\ H_y = -\partial\psi/\partial y \end{cases}$$ | Flux Density: $B_x$, $B_y$ $$B = \nabla \times A \to \begin{cases} B_x = \partial A_z/\partial y \\ B_y = -\partial A_z/\partial x \end{cases}$$ |
| *Potentials and Fields* | *Region I* | $\psi_n^I = \left(A_{1n} e^{nky} + B_{1n} e^{-nky}\right)\sin nkx$ $$B_{xn}^I = -\mu_0 nk\left(A_{1n} e^{nky} + B_{1n} e^{-nky}\right)\cos nkx$$ $$H_{yn}^I = -nk\left(A_{1n} e^{nky} - B_{1n} e^{-nky}\right)\sin nkx$$ | $\psi_n^I = \left(A'_{1n} e^{nky} + B'_{1n} e^{-nky}\right)\sin nkx$ $$H_{xn}^I = -nk\left(A'_{1n} e^{nky} + B'_{1n} e^{-nky}\right)\cos nkx$$ $$H_{yn}^I = -nk\left(A'_{1n} e^{nky} - B'_{1n} e^{-nky}\right)\sin nkx$$ | $A_{zn}^I = \left(A'_{1n} e^{nky} + B'_{1n} e^{-nky}\right)\cos nkx$ $$B_{xn}^I = nk\left(A'_{1n} e^{nky} - B'_{1n} e^{-nky}\right)\cos nkx$$ $$B_{yn}^I = nk\left(A'_{1n} e^{nky} + B'_{1n} e^{-nky}\right)\sin nkx$$ |
| | *Region II* | $\psi_n^{II} = \left(A_{2n} e^{nky} + B_{2n} e^{-nky}\right)\sin nkx$ $$B_{xn}^{II} = -\mu_0 nk\left(A_{2n} e^{nky} + B_{2n} e^{-nky}\right)\cos nkx$$ $$H_{yn}^{II} = -nk\left(A_{2n} e^{nky} - B_{2n} e^{-nky}\right)\sin nkx$$ | $\psi_n^{II} = \left(A'_{2n} e^{nky} + B'_{2n} e^{-nky} + \dfrac{M_{xn}}{nk}\right)\sin nkx$ $$H_{xn}^{II} = -nk\left(A'_{2n} e^{nky} + B'_{2n} e^{-nky} + \dfrac{M_{xn}}{nk}\right)\cos nkx$$ $$H_{yn}^{II} = -nk\left(A'_{2n} e^{nky} - B'_{2n} e^{-nky}\right)\sin nkx$$ | $A_{zn}^{II} = \left(A'_{2n} e^{nky} + B'_{2n} e^{-nky} + \dfrac{\mu_0 M_{yn}}{nk}\right)\cos nkx$ $$B_{xn}^{II} = nk\left(A'_{2n} e^{nky} - B'_{2n} e^{-nky}\right)\cos nkx$$ $$B_{yn}^{II} = nk\left(A'_{2n} e^{nky} + B'_{2n} e^{-nky} + \dfrac{\mu_0 M_{yn}}{nk}\right)\sin nkx$$ |
| *Boundary Condition: Motor with Back-Iron* | *Stator* | Zero: $y = 0 \to B_x^I = 0$ | Zero: $y = 0 \to H_x^I = 0$ | Zero: $y = 0 \to B_x^I = 0$ |
| | *y-Field* | Discontinuity = $\sigma_m$ $$\nabla \cdot \mu_0 \vec{H} = \rho_m$$ $g = g_e \to H_y^{II} - H_y^I = \sigma_m / \mu_0$ | Discontinuity = $M_y$ $$\nabla \cdot B = 0 \to \nabla \cdot H = -\nabla \cdot M$$ $g = g_e \to \mu_0 \overset{B_y^I}{\overbrace{H_y^I}} = \overset{B_y^{II}}{\overbrace{\mu_0(H_y^{II} + M_y)}}$ | No discontinuity $$\nabla \cdot B = 0$$ $g = g_e \to B_y^I = B_y^{II}$ |
| | *x-Field* | Discontinuity = $K_m$ $$\nabla \times (B / \mu_0) = J_m$$ $g = g_e \to B_x^I - B_x^{II} = \mu_0 K_m$ $g = g_e + h_m \to B_x^{II} = -\mu_0 K_m$ | $\to H_x^I = H_x^{II} = M_x$ No discontinuity $$\nabla \times H = 0$$ $g = g_e \to H_x^I = H_x^{II}$ $g = g_e + h_m \to H_x^{II} = H_x^{iron} = 0$ | Discontinuity = $M_x$ $$\nabla \times H = 0 \to \nabla \times B = \mu_0 \nabla \times M$$ $g = g_e \to \overset{H_x^I}{\overbrace{\dfrac{B_x^I}{\mu_0}}} = \overset{H_x^{II}}{\overbrace{\dfrac{B_x^{II}}{\mu_0} - M_x}}$ $\to B_x^I - B_x^{II} = -\mu_0 M_x$ $g = g_e + h_m \to \overset{H_x^{II}}{\overbrace{\dfrac{B_x^{II}}{\mu_0} - M_x}} = 0$ $\to B_x^{II} = \mu_0 M_x$ |

$$\psi_n = \left(A_n e^{nky} + B_n e^{-nky} + \dfrac{M_{xn}}{nk}\right)\sin nkx \qquad (A21)$$

First, the solutions and the boundary conditions need to be written in terms of field intensity $H$, and at the end flux density $B$ is obtained.

### A. The Motor without Back-Iron

For the motor without back-iron behind PMs, the general solutions for the three regions are derived as follows

4. Region I: coil and airgap

An odd function of scalar potential can be expressed as in below:

$$\psi^I = \sum_{n=1,3,\dots}^{+\infty} \psi_n^I = \sum_{n=1,3,\dots}^{+\infty} \overset{\psi_n^I}{\overbrace{\left(A'_{1n} e^{nky} + B'_{1n} e^{-nky}\right)\sin nkx}} \qquad (A22)$$

Then, field intensity can be obtained as follows:

$$H_x^I = \sum_{n=1,3,\dots}^{+\infty} H_{xn}^I = \sum_{n=1,3,\dots}^{+\infty} \overbrace{-nk\left(A_{1n}' e^{nky} + B_{1n}' e^{-nky}\right)\cos nkx}^{H_{\infty}^I} \quad (A23)$$

$$H_y^I = \sum_{n=1,3,\dots}^{+\infty} H_{yn}^I = \sum_{n=1,3,\dots}^{+\infty} \overbrace{-nk\left(A_{1n}' e^{nky} - B_{1n}' e^{-nky}\right)\sin nkx}^{H_{\infty}^I}$$

Using the boundary condition at the surface of the stator ($A_{1n}=-B_{1n}$) that will be discussed later, it can be simplified to:

$$H_x^I = \sum_{n=1,3,\dots}^{+\infty} H_{xn}^I = \sum_{n=1,3,\dots}^{+\infty} \overbrace{-2nk\,A_{1n}' \sinh nky\,\cos nkx}^{H_{\infty}^I} \quad (A24)$$

$$H_y^I = \sum_{n=1,3,\dots}^{+\infty} H_{yn}^I = \sum_{n=1,3,\dots}^{+\infty} \overbrace{-2nk\,A_{1n}' \cosh nky\,\sin nkx}^{H_{\infty}^I}$$

5. Region II: Halbach array

The general solution for the scalar potential in the Halbach array region, which includes a particular solution, can be written as in below:

$$\psi^{II} = \sum_{n=1,3,\dots}^{+\infty} \psi_n^{II} = \sum_{n=1,3,\dots}^{+\infty} \overbrace{\left(A_{2n}' e^{nky} + B_{2n}' e^{-nky} + \frac{M_{xn}}{nk}\right)\sin nkx}^{\psi_n^{II}} \quad (A25)$$

Then, field intensity can be expressed as follows:

$$H_x^{II} = \sum_{n=1,3,\dots}^{+\infty} H_{xn}^{II} = \sum_{n=1,3,\dots}^{+\infty} \overbrace{-nk\left(A_{2n}' e^{nky} + B_{2n}' e^{-nky} + \frac{M_{xn}}{nk}\right)\cos nkx}^{H_{\infty}^{II}}$$

$$H_y^{II} = \sum_{n=1,3,\dots}^{+\infty} H_{yn}^{II} = \sum_{n=1,3,\dots}^{+\infty} \overbrace{-nk\left(A_{2n}' e^{nky} - B_{2n}' e^{-nky}\right)\sin nkx}^{H_{\infty}^{II}}$$

$$(A26)$$

6. Region III: free space behind PMs

Requiring the potentials and magnetic fields to be finite at $y=+\infty$, the coefficient $A_{3n}$ must be zero. Consequently, the general solution for the scalar potential is expressed as follows:

$$\psi^{III} = \sum_{n=1,3,\dots}^{+\infty} \psi_n^{III} = \sum_{n=1,3,\dots}^{+\infty} \overbrace{B_{3n}' e^{-nky} \sin nkx}^{\psi_n^{III}} \quad (A27)$$

Then, the field intensities are obtained as in below:

$$H_x^{III} = \sum_{n=1,3,\dots}^{+\infty} H_{xn}^{III} = \sum_{n=1,3,\dots}^{+\infty} \overbrace{-nk\,B_{3n}' e^{-nky}\cos nkx}^{H_{\infty}^{III}}$$

$$H_y^{III} = \sum_{n=1,3,\dots}^{+\infty} H_{yn}^{III} = \sum_{n=1,3,\dots}^{+\infty} \overbrace{nk\,B_{3n}' e^{-nky}\sin nkx}^{H_{\infty}^{III}}$$

$$(A28)$$

To determine the values of the above six unknow ns, a set of six boundary condition s is needed to create a six-by-six system of equations as in the following:

1. Stator surface ($H_x$ at $y=0$)

$$\nabla \times H = 0 \;\Rightarrow\; H_x^I\big|_{y=0} = H_x^{iron} = 0 \;\Rightarrow\; A_{1n}' + B_{1n}' = 0 \quad (A29)$$

2. Bottom of Halbach array ($B_y$ at $y=g_e$)

$$\nabla . B = 0 \Rightarrow B_y^I = B_y^{II} \Rightarrow \mu_0 H_y^I = \mu_0 (H_y^{II} + M_y) \quad (A30)$$

$$A_{1n}' e^{nkg_e} - B_{1n}' e^{-nkg_e} - A_{2n}' e^{nkg_e} + B_{2n}' e^{-nkg_e} = -\frac{M_y}{nk} \quad (A31)$$

3. Bottom of Halbach array ($H_x$ at $y=g_e$)

$$\nabla \times H = 0 \Rightarrow H_x^I = H_x^{II} \quad (A32)$$

$$-A_{1n}' e^{nkg_e} - B_{1n}' e^{-nkg_e} + A_{2n}' e^{nkg_e} + B_{2n}' e^{-nkg_e} = -\frac{M_{xn}}{nk} \quad (A33)$$

4. Top of Halbach array ($B_y$ at $y=g_e+h_m$)

$$\nabla . B = 0 \Rightarrow B_y^{II} = B_y^{III} \Rightarrow \mu_0 (H_y^{II} + M_y) = \mu_0 H_y^{III} \quad (A34)$$

$$A_{2n}' e^{nk(g_e+h_m)} - B_{2n}' e^{-nk(g_e+h_m)} + B_{3n}' e^{-nk(g_e+h_m)} = \frac{M_{yn}}{nk} \quad (A35)$$

5. Top of Halbach array ($H_x$ at $y=g_e+h_m$)

$$\nabla \times H = 0 \Rightarrow H_x^{II} = H_x^{III} \quad (A36)$$

$$-A_{2n}' e^{nk(g_e+h_m)} - B_{2n}' e^{-nk(g_e+h_m)} + B_{3n}' e^{-nk(g_e+h_m)} = \frac{M_{xn}}{nk} \quad (A37)$$

6. Infinity ($y=+\infty$)

$$A_{3n}' = 0 \quad (A38)$$

The aforementioned equations lead to the following 5-by-5 system of equations whose solution is given in the Appendix.

$$\begin{bmatrix} 1 & 1 & 0 & 0 & 0 \\ e^{nkg_e} & -e^{-nkg_e} & -e^{nkg_e} & e^{-nkg_e} & 0 \\ -e^{nkg_e} & -e^{-nkg_e} & e^{nkg_e} & e^{-nkg_e} & 0 \\ 0 & 0 & e^{nk(g_e+h_m)} & -e^{-nk(g_e+h_m)} & e^{-nk(g_e+h_m)} \\ 0 & 0 & -e^{nk(g_e+h_m)} & -e^{-nk(g_e+h_m)} & e^{-nk(g_e+h_m)} \end{bmatrix} \begin{bmatrix} A_{1n}' \\ B_{1n}' \\ A_{2n}' \\ B_{2n}' \\ B_{3n}' \end{bmatrix} = \begin{bmatrix} 0 \\ -M_y/nk \\ -M_{xn}/nk \\ M_y/nk \\ M_{xn}/nk \end{bmatrix}$$

$$(A39)$$

Solving this system of equation yields the six unknowns as follows:

$$A_{1n}' = A_{1n} \;;\; A_{2n}' = A_{2n} \quad (A40)$$

$$B_{1n}' = B_{1n} \;;\; B_{2n}' = B_{2n} \;;\; B_{3n}' = B_{3n} \quad (A41)$$

### B. The Motor with Back-Iron

For the motor with back-iron behind PMs, the general solutions for the three regions are derived as follows:

1. Region I: coil and airgap

An odd function of scalar potential can be expressed as in below:

$$\psi^I = \sum_{n=1,3,\dots}^{+\infty} \psi_n^I = \sum_{n=1,3,\dots}^{+\infty} \overbrace{\left(A_{1n}' e^{nky} + B_{1n}' e^{-nky}\right)\sin nkx}^{\psi_n^I} \quad (A42)$$

Then, field intensity can be expressed as follows:

$$H_x^I = \sum_{n=1,3,\dots}^{+\infty} H_{xn}^I = \sum_{n=1,3,\dots}^{+\infty} \overbrace{-nk\left(A_{1n}' e^{nky} + B_{1n}' e^{-nky}\right)\cos nkx}^{H_{\infty}^I} \quad (A43)$$

$$H_y^I = \sum_{n=1,3,\dots}^{+\infty} H_{yn}^I = \sum_{n=1,3,\dots}^{+\infty} \overbrace{-nk\left(A_{1n}' e^{nky} - B_{1n}' e^{-nky}\right)\sin nkx}^{H_{\infty}^I}$$

Using the boundary condition at the surface of the stator ($A_{1n}=-B_{1n}$) that will be discussed later, it can be simplified to:

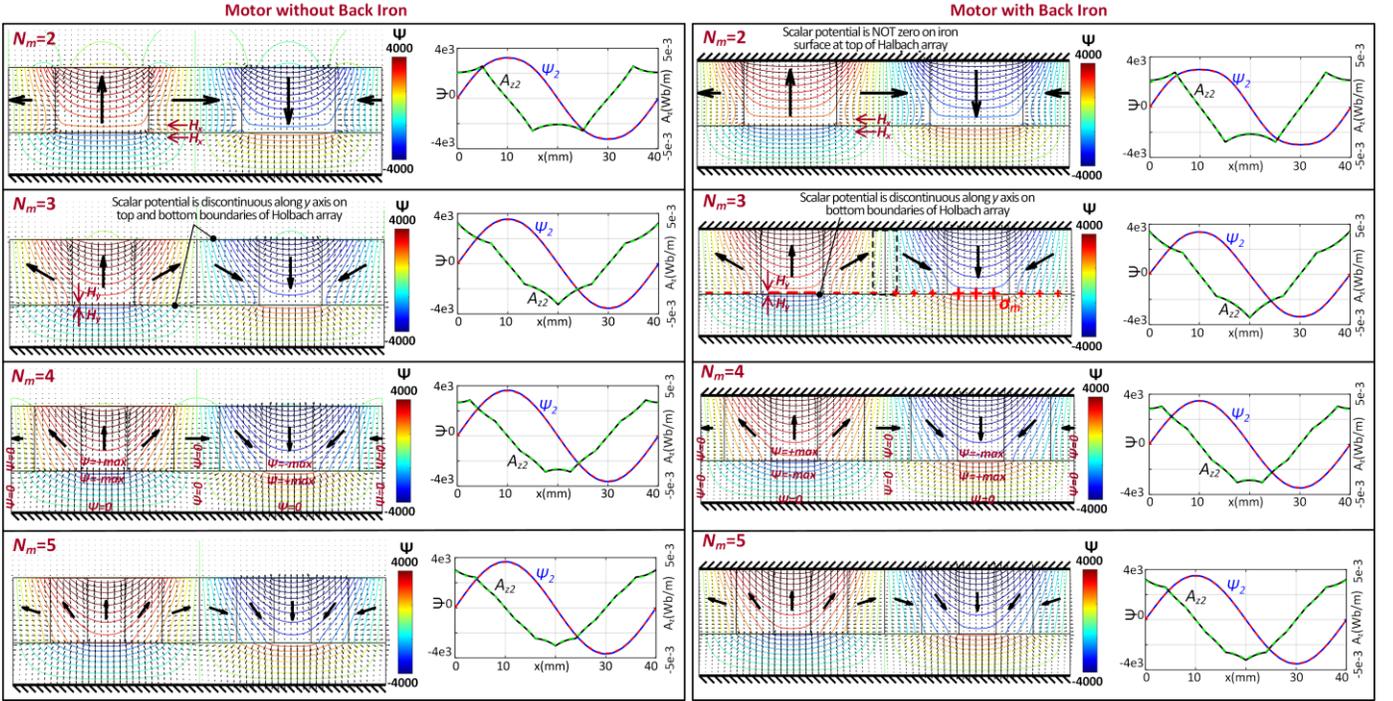

Fig. 15. Model 1: distribution of scalar equipotential lines along with field intensity vectors, and potential distributions within the center of Halbach array $y=g_e+h_m$ (solid lines: model; dashed lines: FEM)

$$\begin{cases} H_x^I = \sum_{n=1,3,...}^{+\infty} H_{xn}^I = \sum_{n=1,3,...}^{+\infty} \overbrace{-2nk\,A'_{1n}\sinh nky\,\cos nkx}^{H_{xn}^I} \\ H_y^I = \sum_{n=1,3,...}^{+\infty} H_{yn}^I = \sum_{n=1,3,...}^{+\infty} \overbrace{-2nk\,A'_{1n}\cosh nky\,\sin nkx}^{H_{yn}^I} \end{cases} \quad \text{(A44)}$$

2. Region II: Halbach array

The general solution for the scalar potential is as follows:

$$\psi^{II} = \sum_{n=1,3,...}^{+\infty} \psi_n^{II} = \sum_{n=1,3,...}^{+\infty} \overbrace{\left(A'_{2n}\,e^{nky} + B'_{2n}\,e^{-nky} + \frac{M_{xn}}{nk}\right)\sin nkx}^{\psi_x^{II}} \quad \text{(A45)}$$

Then, field intensity can be expressed as follows:

$$\begin{cases} H_x^{II} = \sum_{n=1,3,...}^{+\infty} H_{xn}^{II} = \sum_{n=1,3,...}^{+\infty} \overbrace{-nk\left(A'_{2n}\,e^{nky} + B'_{2n}\,e^{-nky} + \frac{M_{xn}}{nk}\right)\cos nkx}^{H_{xn}^{II}} \\ H_y^{II} = \sum_{n=1,3,...}^{+\infty} H_{yn}^{II} = \sum_{n=1,3,...}^{+\infty} \overbrace{-nk\left(A'_{2n}\,e^{nky} - B'_{2n}\,e^{-nky}\right)\sin nkx}^{H_{yn}^{II}} \end{cases} \quad \text{(A46)}$$

To determine the values of the above four unknowns, a set of six boundary conditions is needed to create a four-by-four system of equations as in the following:

1. Stator surface ($H_x$ at y=0)

$$\nabla \times H = 0 \;\Rightarrow H_x^I\big|_{y=0} = H_x^{iron} = 0 \Rightarrow A'_{1n} + B'_{1n} = 0 \quad \text{(A47)}$$

2. Bottom of Halbach array ($B_y$ at $y=g_e$)

$$\nabla . B = 0 \Rightarrow B_y^I = B_y^{II} \Rightarrow \mu_0 H_y^I = \mu_0(H_y^{II} + M_y) \quad \text{(A48)}$$

$$A'_{1n}\,e^{nkg_e} - B'_{1n}\,e^{-nkg_e} - A'_{2n}\,e^{nkg_e} + B'_{2n}\,e^{-nkg_e} = -\frac{M_y}{nk} \quad \text{(A49)}$$

3. Bottom of Halbach array ($H_x$ at y=$g_e$)

$$\nabla \times H = 0 \Rightarrow H_x^I = H_x^{II} \quad \text{(A50)}$$

$$-A'_{1n}\,e^{nkg_e} - B'_{1n}\,e^{-nkg_e} + A'_{2n}\,e^{nkg_e} + B'_{2n}\,e^{-nkg_e} = -\frac{M_{xn}}{nk} \quad \text{(A51)}$$

4. Top of Halbach array ($H_x$ at $y=g_e+h_m$)

$$\nabla \times H = 0 \Rightarrow H_x^{II} = H_x^{iron} = 0 \quad \text{(A52)}$$

$$-A'_{2n}\,e^{nk(g_e+h_m)} - B'_{2n}\,e^{-nk(g_e+h_m)} = \frac{M_{xn}}{nk} \quad \text{(A53)}$$

The aforementioned equations lead to the following 4-by-4 system of equations whose solution is given in the Appendix.

$$\begin{bmatrix} 1 & 1 & 0 & 0 \\ e^{-nkg_e} & -e^{-nkg_e} & -e^{-nkg_e} & -e^{-nkg_e} \\ -e^{-nkg_e} & -e^{-nkg_e} & e^{nkg_e} & e^{-nkg_e} \\ 0 & 0 & -e^{nk(g_e+h_m)} & -e^{-nk(g_e+h_m)} \end{bmatrix} \begin{bmatrix} A'_{1n} \\ B'_{1n} \\ A'_{2n} \\ B'_{2n} \end{bmatrix} = \begin{bmatrix} 0 \\ -M_y/nk \\ -M_{xn}/nk \\ M_{xn}/nk \end{bmatrix} \quad \text{(A54)}$$

$$A'_{1n} = A_{1n}\;;\; A'_{2n} = A_{2n} \quad \text{(A55)}$$

$$B'_{1n} = B_{1n}\;;\; B'_{2n} = B_{2n} \quad \text{(A56)}$$

## Appendix IV: Field Solution Using Poisson's Equation in Terms of Vector Magnetic Potential (Model 3)

Field solutions can also be obtained using Poisson's equation in terms of the magnetic vector potential. Divergence of the curl of any vector field is zero. Thus, according to Gausses' law, a magnetic vector potential $A$ can be defined as follows:

$$\nabla . B = 0 \to B = \nabla \times A \quad \text{(A57)}$$

Substituting the identity $\nabla \times \nabla \times A = \nabla(\nabla . A) - \nabla^2 A$ into

$$\nabla \times H = J \xrightarrow{H = B/\mu_0 - M} \nabla \times \left( \frac{\nabla \times A}{\mu_0} - M \right) = J \quad (A58)$$

$$\rightarrow \nabla^2 A - \nabla(\nabla . A) = -\mu_0 \left( J + \nabla \times M \right)$$

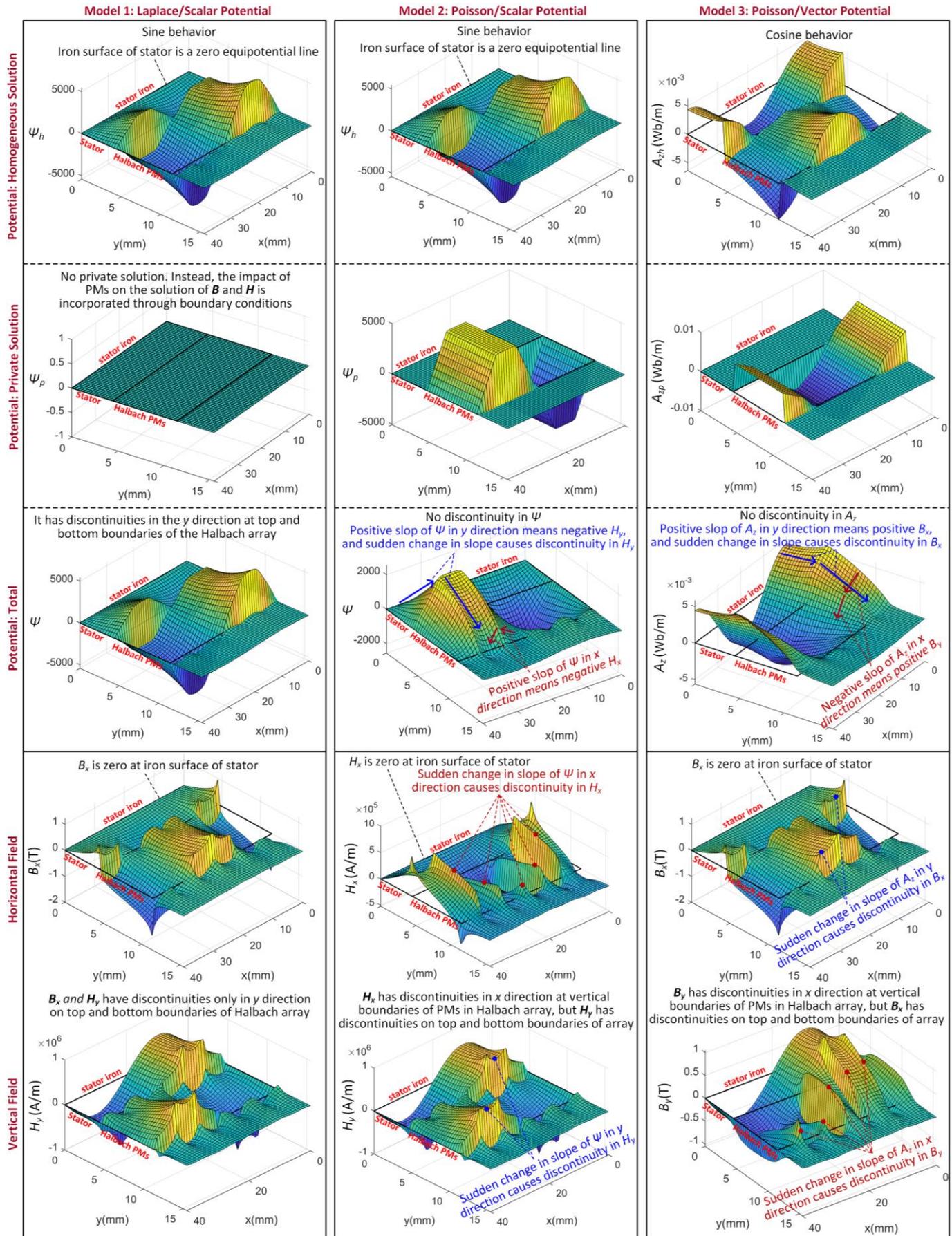

Fig. 16. Two-dimensional distribution of magnetic potentials and fields for the motor without back-iron behind PMs ($N_m$=3).

Using Coulomb's gauge condition $\nabla . A = 0$, Poisson's equation is obtained. It is a 2D problem with $z$-directed currents and vector potentials as in below:

$$\nabla^2 A_z = -\mu_0 \left( J_z + J_M \right) \; ; \; J_M = \nabla \times M \quad (A59)$$

It requires a homogeneous solution $A_{z,h}$ and a particular solution $A_{z,p}$ as in below:

$$A_z = A_{z,h} + A_{z,p} \quad (A60)$$

The homogeneous solution solves the Laplacian equation by employing *the separation of variables* as follows:

$$\nabla^2 A_{z,h} = 0 \; ; \; A_{z,h}(x,y) = X'(x) Y'(y) \quad (A61)$$

where the general solutions are obtained as follows:

$$\begin{cases} X'(x) \sim e^{jkx}, \; e^{-jkx} \quad or \quad \sin kx, \; \cos kx \\ Y'(y) \sim e^{ky}, \; e^{-ky} \quad or \quad \sinh ky, \; \cosh ky \end{cases} \quad (A62)$$

As the behavior of vector potential is similar to currents, here Amperian currents, the homogeneous solution for all three regions is an even function of $x$ as follows:

$$A_{zn,h} = \left( A_n e^{nky} + B_n e^{-nky} \right) \cos nkx \quad (A63)$$

The particular solution is zero except within the Halbach array, where we need to guess a solution that satisfies the Poisson's equation:

$$\nabla^2 A_{z,p} = -\mu_0 \nabla \times M = -\mu_0 \left( \frac{\partial M_y}{\partial x} - \frac{\partial M_x}{\partial y} \right) \quad (A64)$$

It leads to:

$$\frac{\partial^2 A_{z,p}}{\partial x^2} + \frac{\partial^2 A_{z,p}}{\partial y^2} = \sum_{n=1,3,\dots}^{+\infty} -\mu_0 nk M_{yn} \cos nkx \quad (A65)$$

Particular solution is not unique, and any solution satisfying the above equation works. Assuming a particular solution that is only a function of $x$, after mathematical manipulations, we can guess the following solution that works:

$$A_{z,p} = \sum_{n=1,3,\dots}^{+\infty} A_{z,pn} = \sum_{n=1,3,\dots}^{+\infty} \frac{\mu_0 M_{yn}}{nk} \cos nkx \quad (A66)$$

Then, the general solution for Halbach array region is obtained as follows:

$$A_z = \sum_{n=1,3,\dots}^{+\infty} A_{zn} = \sum_{n=1,3,\dots}^{+\infty} \left( A_n e^{nky} + B_n e^{-nky} + \frac{\mu_0 M_{yn}}{nk} \right) \cos nkx \quad (A67)$$

At the end, the solutions and the boundary conditions need to be written in terms of flux density $B$ as follows:

$$\begin{cases} B_x = \frac{\partial A_z}{\partial y} \Rightarrow B_{xn} = \frac{\partial A_{zn}}{\partial y} \\ B_y = -\frac{\partial A_z}{\partial x} \Rightarrow B_{yn} = -\frac{\partial A_{zn}}{\partial x} \end{cases} \quad (A68)$$

The above equation tells us that flux density vectors are tangential to equipotential lines of vector potential $A_z$.

### C. The Motor without Back-Iron

For the motor without back-iron behind PMs, the general solutions for the three regions are derived as follows:

#### 1. Region I: coil and airgap

The general solution of vector potential can be expressed as in below:

$$A_z^I = \sum_{n=1,3,\dots}^{+\infty} A_{zn}^I = \sum_{n=1,3,\dots}^{+\infty} \overbrace{\left( A_{1n}' e^{nky} + B_{1n}' e^{-nky} \right) \cos nkx}^{A_{zn}^I} \quad (A69)$$

Then, flux density can be expressed as follows:

$$\begin{cases} B_x^I = \sum_{n=1,3,\dots}^{+\infty} B_{xn}^I = \sum_{n=1,3,\dots}^{+\infty} \overbrace{nk \left( A_{1n}' e^{nky} - B_{1n}' e^{-nky} \right) \cos nkx}^{B_{xn}^I} \\ B_y^I = \sum_{n=1,3,\dots}^{+\infty} B_{yn}^I = \sum_{n=1,3,\dots}^{+\infty} \overbrace{nk \left( A_{1n}' e^{nky} + B_{1n}' e^{-nky} \right) \sin nkx}^{B_{yn}^I} \end{cases} \quad (A70)$$

Using the boundary condition at the surface of the stator ($A_{1n} = B_{1n}$), it can be simplified to:

$$\begin{cases} B_x^I = \sum_{n=1,3,\dots}^{+\infty} B_{xn}^I = \sum_{n=1,3,\dots}^{+\infty} \overbrace{2nk \, A_{1n}' \sinh nky \, \cos nkx}^{B_{xn}^I} \\ B_y^I = \sum_{n=1,3,\dots}^{+\infty} B_{yn}^I = \sum_{n=1,3,\dots}^{+\infty} \overbrace{2nk \, A_{1n}' \cosh nky \, \sin nkx}^{B_{yn}^I} \end{cases} \quad (A71)$$

#### 2. Region II: Halbach array

The general solution for the vector potential is as follows:

$$A_z^{II} = \sum_{n=1,3,\dots}^{+\infty} A_{zn}^{II} = \sum_{n=1,3,\dots}^{+\infty} \overbrace{\left( A_{2n}' e^{nky} + B_{2n}' e^{-nky} + \frac{\mu_0 M_{yn}}{nk} \right) \cos nkx}^{A_{zn}^{II}} \quad (A72)$$

Then, flux density can be obtained as follows:

$$\begin{cases} B_x^{II} = \sum_{n=1,3,\dots}^{+\infty} B_{xn}^{II} = \sum_{n=1,3,\dots}^{+\infty} \overbrace{nk \left( A_{2n}' e^{nky} - B_{2n}' e^{-nky} \right) \cos nkx}^{B_{xn}^{II}} \\ B_y^{II} = \sum_{n=1,3,\dots}^{+\infty} B_{yn}^{II} = \sum_{n=1,3,\dots}^{+\infty} \overbrace{nk \left( A_{2n}' e^{nky} + B_{2n}' e^{-nky} + \frac{\mu_0 M_{yn}}{nk} \right) \sin nkx}^{B_{yn}^{II}} \end{cases} \quad (A73)$$

#### 3. Region III: free space behind PMs

The general solution for vector potentials is as follows:

$$A_z^{III} = \sum_{n=1,3,\dots}^{+\infty} A_{zn}^{III} = \sum_{n=1,3,\dots}^{+\infty} \overbrace{\left( A_{3n}' e^{nky} + B_{3n}' e^{-nky} \right) \cos nkx}^{A_{zn}^{III}} \quad (A74)$$

Requiring the coefficient $A_{3n}$ to be zero at infinity, flux density components are obtained as in below:

$$\begin{cases} B_x^{III} = \sum_{n=1,3,\dots}^{+\infty} B_{xn}^{III} = \sum_{n=1,3,\dots}^{+\infty} \overbrace{nk \left( A_{3n}' e^{nky} - B_{3n}' e^{-nky} \right) \cos nkx}^{B_{xn}^{III}} \\ B_y^{III} = \sum_{n=1,3,\dots}^{+\infty} B_{yn}^{III} = \sum_{n=1,3,\dots}^{+\infty} \overbrace{nk \left( A_{3n}' e^{nky} + B_{3n}' e^{-nky} \right) \sin nkx}^{B_{yn}^{III}} \end{cases} \quad (A75)$$

To determine the values of the above six unknowns, a set of six boundary conditions is needed to create a six-by-six system of equations as in the following:

#### 1. Stator surface ($H_x$ at $y=0$)

$$\nabla \times H = 0 \Rightarrow H_x^I = H_x^{iron} = 0 \Rightarrow -A_{1n}' + B_{1n}' = 0 \quad (A76)$$

2. Bottom of Halbach array ($B_y$ at $y=g_e$)

$$\nabla . B = 0 \Rightarrow B_y^I = B_y^{II} \tag{A77}$$

$$-A_{1n}' e^{nkg_e} - B_{1n}' e^{-nkg_e} + A_{2n}' e^{nkg_e} + B_{2n}' e^{-nkg_e} = -\frac{\mu_0 M_{yn}}{nk} \tag{A78}$$

3. Bottom of Halbach array ($H_x$ at $y=g_e$)

$$\nabla \times H = 0 \Rightarrow H_x^I = H_x^{II} \Rightarrow \frac{B_x^I}{\mu_0} = \frac{B_x^{II}}{\mu_0} - M_x \tag{A79}$$

$$A_{1n}' e^{nkg_e} - B_{1n}' e^{-nkg_e} - A_{2n}' e^{nkg_e} + B_{2n}' e^{-nkg_e} = -\frac{\mu_0 M_{xn}}{nk} \tag{A80}$$

4. Top of Halbach array ($B_y$ at $y=g_e+h_m$)

$$\nabla . B = 0 \Rightarrow B_y^{II} = B_y^{III} \tag{A81}$$

$$-A_{2n}' e^{nk(g_e+h_m)} - B_{2n}' e^{-nk(g_e+h_m)} + B_{3n}' e^{-nk(g_e+h_m)} = \frac{\mu_0 M_{yn}}{nk} \tag{A82}$$

5. Top of Halbach array ($H_x$ at $y=g_e+h_m$)

$$\nabla \times H = 0 \Rightarrow H_x^{II} = H_x^{III} \Rightarrow \frac{B_x^{II}}{\mu_0} - M_x = \frac{B_x^{III}}{\mu_0} \tag{A83}$$

$$A_{2n}' e^{nk(g_e+h_m)} - B_{2n}' e^{-nk(g_e+h_m)} + B_{3n}' e^{-nk(g_e+h_m)} = \frac{\mu_0 M_{xn}}{nk} \tag{A84}$$

6. Infinity ($y=+\infty$)

$$A_{3n}' = 0 \tag{A85}$$

The aforementioned equations lead to the following 5-by-5 system of equations whose solution is given in the Appendix.

$$\begin{bmatrix} -1 & 1 & 1 & 0 & 0 \\ -e^{-nkg_e} & -e^{-nkg_e} & e^{nkg_e} & e^{-nkg_e} & 0 \\ e^{nkg_e} & -e^{-nkg_e} & -e^{nkg_e} & e^{-nkg_e} & 0 \\ 0 & 0 & -e^{nk(g_e+h_m)} & -e^{-nk(g_e+h_m)} & e^{nk(g_e+h_m)} \\ 0 & 0 & e^{nk(g_e+h_m)} & -e^{-nk(g_e+h_m)} & -e^{-nk(g_e+h_m)} \end{bmatrix} \begin{bmatrix} A_{1n}' \\ B_{1n}' \\ A_{2n}' \\ B_{2n}' \\ B_{3n}' \end{bmatrix} = \begin{bmatrix} 0 \\ -\mu_0 M_{yn}/nk \\ -\mu_0 M_{xn}/nk \\ \mu_0 M_{yn}/nk \\ \mu_0 M_{xn}/nk \end{bmatrix} \tag{A86}$$

Solving this system of equation yields the unknowns as follows:

$$A_{1n}' = -\mu_0 A_{1n} ; \ A_{2n}' = -\mu_0 A_{2n} \tag{A87}$$

$$B_{1n}' = \mu_0 B_{1n} ; \ B_{2n}' = \mu_0 B_{2n} ; \ B_{3n}' = \mu_0 B_{3n} \tag{A88}$$

### D. The Motor with Back-Iron

For the motor with back-iron behind PMs, the general solutions for the three regions are derived as follows:

#### 1. Region I: coil and airgap

The general solution of vector potential can be expressed as in below:

$$A_z^I = \sum_{n=1,3,...}^{+\infty} A_{zn}^I = \sum_{n=1,3,...}^{+\infty} \overbrace{\left( A_{1n}' e^{nky} + B_{1n}' e^{-nky} \right) \cos nkx}^{A_{zn}^I} \tag{A89}$$

Then, the magnetic fields can be obtained as:

$$\begin{cases} B_x^I = \sum_{n=1,3,...}^{+\infty} B_{xn}^I = \sum_{n=1,3,...}^{+\infty} \overbrace{nk\left( A_{1n}' e^{nky} - B_{1n}' e^{-nky} \right) \cos nkx}^{B_{xn}^I} \\ B_y^I = \sum_{n=1,3,...}^{+\infty} B_{yn}^I = \sum_{n=1,3,...}^{+\infty} \overbrace{nk\left( A_{1n}' e^{nky} + B_{1n}' e^{-nky} \right) \sin nkx}^{B_{yn}^I} \end{cases} \tag{A90}$$

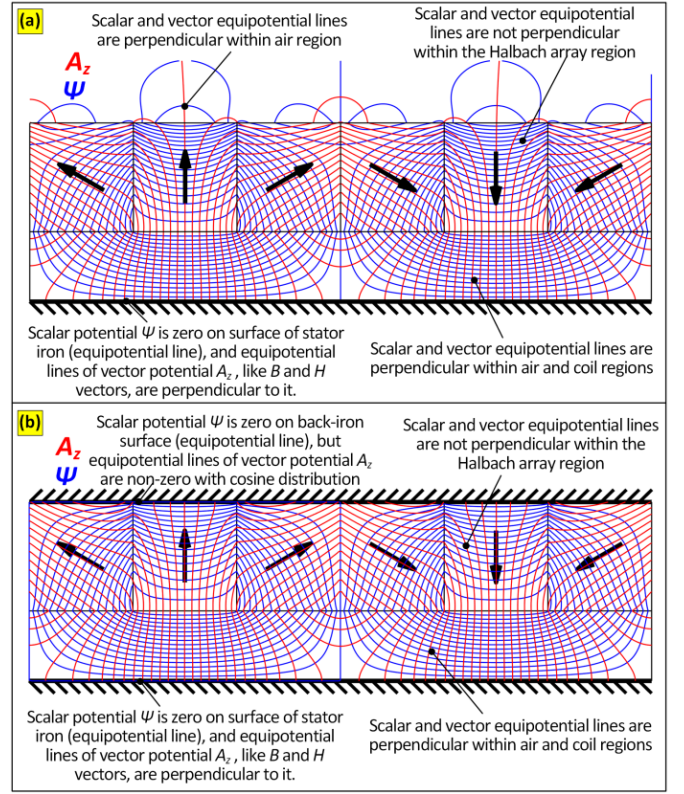

Fig. 17. Scalar and vector equipotential lines for motors (a) without and (b) with back-iron behind PMs.

Using the boundary condition at the surface of the stator ($A_{1n}=B_{1n}$), it can be simplified to:

$$\begin{cases} B_x^I = \sum_{n=1,3,...}^{+\infty} B_{xn}^I = \sum_{n=1,3,...}^{+\infty} \overbrace{2nk \, A_{1n}' \sinh nky \, \cos nkx}^{B_{xn}^I} \\ B_y^I = \sum_{n=1,3,...}^{+\infty} B_{yn}^I = \sum_{n=1,3,...}^{+\infty} \overbrace{2nk \, A_{1n}' \cosh nky \, \sin nkx}^{B_{yn}^I} \end{cases} \tag{A91}$$

#### 2. Region II: Halbach array

The general solution for the vector potential is as follows:

$$A_z^{II} = \sum_{n=1,3,...}^{+\infty} A_{zn}^{II} = \sum_{n=1,3,...}^{+\infty} \overbrace{\left( A_{2n}' e^{nky} + B_{2n}' e^{-nky} + \frac{\mu_0 M_{yn}}{nk} \right) \cos nkx}^{A_{zn}^{II}} \tag{A92}$$

Then the magnetic fields can be expressed as follows:

$$\begin{cases} B_x^{II} = \sum_{n=1,3,...}^{+\infty} B_{xn}^{II} = \sum_{n=1,3,...}^{+\infty} \overbrace{nk\left( A_{2n}' e^{nky} - B_{2n}' e^{-nky} \right) \cos nkx}^{B_{xn}^{II}} \\ B_y^{II} = \sum_{n=1,3,...}^{+\infty} B_{yn}^{II} = \sum_{n=1,3,...}^{+\infty} \overbrace{nk\left( A_{2n}' e^{nky} + B_{2n}' e^{-nky} + \frac{\mu_0 M_{yn}}{nk} \right) \sin nkx}^{B_{yn}^{II}} \end{cases} \tag{A93}$$

To determine the values of the above four unknowns, a set of six boundary conditions is needed to create a four-by-four system of equations as in the following:

#### 1. Stator surface ($H_x$ at $y=0$)

$$\nabla \times H = 0 \Rightarrow H_x^I = H_x^{iron} = 0 \Rightarrow -A_{1n}' + B_{1n}' = 0 \tag{A94}$$

2. Bottom of Halbach array ($B_y$ at $y=g_e$)

$$\nabla . B = 0 \Rightarrow B_y^I = B_y^{II} \tag{A95}$$

$$-A'_{1n} e^{nkg_e} - B'_{1n} e^{-nkg_e} + A'_{2n} e^{nkg_e} + B'_{2n} e^{-nkg_e} = -\frac{\mu_0 M_{yn}}{nk} \tag{A96}$$

3. Bottom of Halbach array ($H_x$ at $y=g_e$)

$$\nabla \times H = 0 \Rightarrow H_x^I = H_x^{II} \Rightarrow \frac{B_x^I}{\mu_0} = \frac{B_x^{II}}{\mu_0} - M_x \tag{A97}$$

$$A'_{1n} e^{nkg_e} - B'_{1n} e^{-nkg_e} - A'_{2n} e^{nkg_e} + B'_{2n} e^{-nkg_e} = -\frac{\mu_0 M_{xn}}{nk} \tag{A98}$$

4. Bottom of Halbach array ($H_x$ at $y=g_e+h_m$)

$$\nabla \times H = 0 \Rightarrow H_x^{II} = H_x^{iron} = 0 \Rightarrow \frac{B_x^{II}}{\mu_0} - M_x = 0 \tag{A99}$$

$$A'_{2n} e^{nk(g_e+h_m)} - B'_{2n} e^{-nk(g_e+h_m)} = \frac{\mu_0 M_{xn}}{nk} \tag{A100}$$

These equations lead to the following 4-by-4 system of equations:

$$\begin{bmatrix} -1 & 1 & 0 & 0 \\ -e^{nkg_e} & -e^{-nkg_e} & e^{nkg_e} & e^{-nkg_e} \\ e^{nkg_e} & -e^{-nkg_e} & -e^{nkg_e} & e^{-nkg_e} \\ 0 & 0 & e^{nk(g_e+h_m)} & -e^{-nk(g_e+h_m)} \end{bmatrix} \begin{bmatrix} A'_{1n} \\ B'_{1n} \\ A'_{2n} \\ B'_{2n} \end{bmatrix} = \begin{bmatrix} 0 \\ -\mu_0 M_{yn}/nk \\ -\mu_0 M_{xn}/nk \\ \mu_0 M_{xn}/nk \end{bmatrix} \tag{A101}$$

Solving this system of equation yields the unknowns as follows:

$$A'_{1n} = -\mu_0 A_{1n} ; A'_{2n} = -\mu_0 A_{2n} \tag{A102}$$

$$B'_{1n} = \mu_0 B_{1n} ; B'_{2n} = \mu_0 B_{2n} \tag{A103}$$

## APPENDIX V: MORE ON FIELD ANALYSIS

This section offers an in-depth analysis of magnetic potentials and fields, providing designers with intuitive insights into the developed analytical models. This enables them to effectively apply the proposed frameworks to different topologies of electromechanical devices and electric machines across various applications.

Figs. 8-11 depict equipotential lines and the distribution of scalar potential derived from Model 2. Fig. 15 shows results obtained from Model 1, while in Fig. 16, for the motor with three PMs per pole and no back-iron, the two-dimensional distribution of homogeneous, particular, and general solutions of scalar and vector potentials, along with field intensity and flux density vectors, are presented for the three developed models. It is observed that the scalar potential in Model 2 and the vector potential in Model 3 are continuous functions of $x$ and $y$, whereas Model 1 exhibits discontinuities on the top and bottom surfaces of the Halbach array. However, derivatives of the scalar potential in Model 1 provide correct solutions for $B_x$ and $H_y$ across the three regions. Specifically, the addition of the particular solution to the homogeneous solution ensures a continuous distribution. Discontinuities in $B_x$, $B_y$, $H_x$, and $H_y$ arise from abrupt changes in the slope of scalar or vector potential in the $x$ or $y$ directions. Notably, $B_x$ and $H_y$ exhibit discontinuities in the $y$-direction at the top and bottom boundaries of the Halbach array, while $B_y$ and $H_x$ show discontinuities in the $x$-direction at the vertical boundaries of the PMs in the Halbach array.

In Fig. 17, equipotential lines of both scalar and magnetic potentials are depicted for motors with and without back-iron behind PMs, showing that they are not necessarily perpendicular. Flux density vectors align tangentially with equipotential lines of vector potential, while field intensity vectors are perpendicular to equipotential lines of scalar potential. In air or coil regions, where $B=\mu_0 H$, equipotential lines of scalar and magnetic potentials are perpendicular because flux density and field intensity vectors are aligned. However, in the Halbach array region where $B=\mu_0(H+M)$, flux density and field intensity vectors are not aligned, so equipotential lines of scalar and magnetic potentials are not perpendicular. The iron surfaces of either the stator or PM back-iron serve as zero equipotential lines. Vector equipotential lines are perpendicular only to the iron surface of the stator.

As summarized in Tables IV and V, it can be compared how magnetization components $M_x$ and $M_y$ are incorporated into the solution:

- In model 1, where there is no particular solution for scalar potential, the impact of vertical magnetization $M_y$ is incorporated into the solution through the boundary conditions of $H_y$ using surface charges $\sigma_m$, and the impact of horizontal magnetization $M_x$ is incorporated into the solution using the boundary conditions of $B_x$ using surface Amperian currents $K_m$.

- In model 2, where there is a particular solution $\Psi_p$ for scalar potential, the impact of vertical magnetization $M_y$ is incorporated into the solution through the boundary conditions of $H_y$ using $M_y$, while the impact of horizontal magnetization $M_x$ is incorporated into solution through the particular solution $\Psi_p$ that will show up in the general solution of $H_x$.

- In model 3, where there is a particular solution $A_{zp}$ for vector potential, the impact of vertical magnetization $M_y$ is incorporated into the solution through the particular solution $A_{zp}$ that will show up in the general solution of $B_y$, while the impact of horizontal magnetization $M_x$ is incorporated into solution through the boundary conditions of $B_x$ using $M_x$.

## APPENDIX VI: SIMPLIFIED EQUATIONS FOR INITIAL SIZING

For initial sizing of the motor, if we assume an approximate normal flux density $B_{av}$ and an average current density $J_{av}$ within the stator area, both uniformly distributed, shear stress is obtained as in below:

$$\tau_{av} = h_c J_{av} B_{av} \tag{A104}$$

Then, the total force for one pole pair pitch is obtained as:

$$F_{av} = \lambda L \tau_{av} \tag{A105}$$

Also, copper loss per unit volume can be approximated as follows:

$$P = \frac{J_{av}^2}{\sigma_{cu}} \tag{A106}$$

**Sajjad Mohammadi** (S'13) received the B.S. degree in electrical engineering from the Kermanshah University of Technology, Kermanshah, Iran, in 2011, the M.S. degree in electrical engineering from the Amirkabir University of Technology, Tehran, Iran, in 2014, and the M.S. and Ph.D. degrees in electrical engineering and computer science from the Massachusetts Institute of Technology (MIT), Cambridge, MA, USA, in 2019 and 2021, respectively. Then, he is now a postdoctoral associate in precision motion control laboratory. His research interests include the design of electric machines, drives, power electronics, and power systems. He attained a number of awards as the 2022 George M. Sprowls Outstanding Ph.D. Thesis Award from MIT, 2014 Best MSc Thesis Award from IEEE Iran Section (nationwide), 2014 Best MSc Thesis Award from Amirkabir University , 2nd place in National Chem-E-Car Competitions 2009 held by Sharif University of Technology, Tehran, 5th place in International Chem-E-Car Competitions 2009 held by McGill University, Montreal, Canada, 2nd place in the Humanoid Robots together with 1st place in Technical Challenge in International Iran-Open 2010 Robotics Competitions, Tehran, 3rd place in National Chem-E-Car Competitions 2010 in Razi University, Kermanshah, and 1st place in the H umanoid Robot League together with the 1st place in Technical Challenge both in International AUTCUP 2010 Khwarizmi Robotics Competitions at Amirkabir University, Tehran.

**James L. Kirtley, Jr.** (M'71–SM'80–F'91–LF'11) received the Ph.D. degree from the Massachusetts Institute of Technology (MIT), Cambridge, MA, USA, in 1971. Currently, he is a Professor of electrical engineering at MIT. He was with General Electric, Large Steam Turbine Generator Department, as an Electrical Engineer, for Satcon Technology Corporation as Vice President and General Manager of the Tech Center and as a Chief Scientist and as the Director. He was Gastdozent at the Swiss Federal Institute of Technology, Zurich (ETH), Switzerland. He is a specialist in electric machinery and electric power systems. Prof. Kirtley, Jr. served as Editor-in-Chief of the IEEE TRANSACTIONS ON ENERGY CONVERSION from 1998 to 2006 and continues to serve as an Editor for that journal and as a member of the Editorial Board of the journal Electric Power Components and Systems. He was awarded the IEEE Third Millennium medal in 2000 and the Nikola Tesla prize in 2002. He was elected to the United States National Academy of Engineering in 2007. He is a Registered Professional Engineer in Massachusetts, USA.

**Jeffrey H. Lang** (F'98) received his SB (1975), SM (1977) and PhD (1980) degrees from the Department of Electrical Engineering and Computer Science at the Massachusetts Institute of Technology. He joined the faculty of MIT in 1980 where he is now the Vitesse Professor of Electrical Engineering. He served as the Associate Director of the MIT Laboratory for Electromagnetic and Electronic Systems from 1991 to 2003, and as an Associate Director of the MIT Microsystems Technology Laboratories from 2012 to 2022. Professor Lang's research and teaching interests focus on the analysis, design and control of electromechanical systems with an emphasis on: rotating machinery; micro/nano-scale (MEMS/NEMS) sensors, actuators and energy converters; flexible structures; and the dual use of electromechanical actuators as motion and force sensors. He has written over 360 papers and holds 36 patents in the areas of electromechanics, MEMS, power electronics and applied control. He has been awarded 6 best-paper prizes from IEEE societies, has received two teaching awards from MIT, and was selected as an MIT MacVicar Fellow in 2022. He is a coauthor of Foundations of Analog and Digital Electronic Circuits published by Morgan Kaufman, and the editor of, and a contributor to, Multi-Wafer Rotating MEMS Machines: Turbines Generators and Engines published by Springer. Professor Lang is a Life Fellow of the IEEE, and a former Hertz

Foundation Fellow. He served as an Associate Editor of Sensors and Actuators between 1991 and 1994. He has also served as the Technical Co-Chair and General Co-Chair of the 1992 and 1993 IEEE MEMS Workshops, respectively, and the General Co-Chair of the 2013 PowerMEMS Conference.

**David L. Trumper** (M' 90) He received the B.S., M.S., and Ph.D. degrees in electrical engineering and computer science from the Massachusetts Institute of Technology, Cambridge, MA, USA, in 1980, 1984, and 1990, respectively. Following the bachelor's degree, he was with the Hewlett-Packard Co. for two years. After finishing the master's degree, he was with the Waters Chromatography Division of Millipore, for two years. Upon completing the Ph.D. degree, he was an Assistant Professor with the Electrical Engineering Department, University of North Carolina at Charlotte, Charlotte, NC, USA, within the precision engineering group. His research centers on the design of precision mechatronic systems, with a focus on the design of novel mechanisms, actuators, sensors, and control systems. His research interests include precision motion control, high-performance manufacturing equipment, novel measurement instruments, biomedical and bioinstrumentation devices, and high-precision magnetic suspensions and bearings. He is a Member of the ASME and ASPE (past-President).